\documentclass{article}

\usepackage{Setup}
\usepackage[numbers]{natbib}
\usepackage{authblk}
\usepackage{empheq}

\usepackage{hyperref}
\usepackage{verbatim}
\usepackage{caption}
\usepackage{mwe}
\usepackage{bm}
\hypersetup{
    colorlinks=true,
    linkcolor=blue,
    filecolor=magenta,      
    urlcolor=cyan,
    pdftitle={Overleaf Example},
    pdfpagemode=FullScreen,
    }

\title{Bayes-optimal inference for \\ spreading processes on random networks}

\author[1]{Davide Ghio}
\author[2]{Antoine L. M. Aragon}
\author[2]{Indaco Biazzo}
\author[2]{Lenka Zdeborová}
\affil[1]{Ecole Polytechnique Fédérale de Lausanne (EPFL). IdePHICS Laboratory}
\affil[2]{Ecole Polytechnique Fédérale de Lausanne (EPFL). SPOC Laboratory}

\begin{document}

\maketitle 
\begin{abstract}
We consider a class of spreading processes on networks, which generalize commonly used epidemic models such as the SIR model or the SIS model with a bounded number of re-infections. We analyse the related problem of inference of the dynamics based on its partial  observations. 
We analyse these inference problems on random networks via a message-passing inference algorithm derived from the Belief Propagation (BP) equations. 
We investigate whether said algorithm solves the problems in a Bayes-optimal way, i.e. no other algorithm can reach a better performance. For this, we leverage the so-called Nishimori conditions that must be satisfied by a Bayes-optimal algorithm. We also probe for phase transitions by considering the convergence time and by initializing the algorithm in both a random and an informed way and comparing the resulting fixed points. We present the corresponding phase diagrams. 
We find large regions of parameters where even for moderate system sizes the BP algorithm converges and satisfies closely the Nishimori conditions, and the problem is thus conjectured to be solved optimally in those regions. 
In other limited areas of the space of parameters, the Nishimori conditions are no longer satisfied and the BP algorithm struggles to converge. No sign of a phase transition is detected, however, and we attribute this failure of optimality to finite-size effects.
The article is accompanied by a Python implementation of the algorithm that is  easy to use or adapt. 
\end{abstract}
\section{\label{sec:intro}Introduction}

Spreading models on graphs are ubiquitous in many scientific domains. Examples range from modelling epidemic spreading processes \cite{Pastor-Satorras2015}, diffusion or relaxing out of equilibrium dynamics in physical systems \cite{henkel2008non}, social dynamics processes, and spreading processes in technological systems \cite{vespignani2012modelling}.  

In inference on spreading  models, we typically want to recover some missing information, having a limited amount of observations on the system. Examples are the identification of the sources of a spreading process (the well-known patient-zero problem) \cite{shelke2019source,2020Castellanos}, epidemic mitigation efforts \cite{EpiMit, InfluEpiMit, brooks2014dynamic}, or the anonymization of Bitcoin transactions \cite{fanti2017deanonymization}.
Inference problems on spreading models have a natural representation in a Bayesian framework.
In the Bayesian approach, the spreading model is the prior, the partial observations on the system induce the likelihood, and the inference problem reduces to the computation of marginals of the posterior distribution. 
We will give a more precise formulation of the Bayesian approach to the problems and the assumptions in Section~\ref{sec:BaySet}.

A large volume of literature in different domains undertakes inference problems on spreading processes. Many are dedicated to the patient-zero problem, where the goal is to infer the nodes (patients) that start the spreading process.  
This problem was introduced in two different application areas \cite{5961801, 10.1103/physrevlett.109.068702}. 
In \cite{altarelli2014patient} the Belief Propagation (BP) equations to solve these inference problems were derived. Nevertheless, regarding their range of applicability and generality, we believe these results are not well shared among communities beyond the statistical physics ones. For instance, in a review on the patient-zero problem \cite{shelke2019source}, the works based on Belief propagation are not mentioned, and plenty of new heuristic algorithms are presented, for instance, \cite{paluch2018fast, yu2017rumor, jain2016fast, cai2018information, jiang2016rumor}, to solve the patient-zero problem on sparse tree-like graphs without comparing the results with the BP approach. 

This is, in particular, striking for two reasons. First, the BP-based inference algorithm obtained better results than competitors across various cases, and graph structures \cite{BIofEpi, braunstein2019network, EpiMit, biazzo2021epidemic}.
Second, on sparse random graphs the BP inference algorithms designed for other problems were argued to obtain an asymptotically exact estimation of the marginals of the posterior and thus the Bayes-optimal inference, see e.g. \cite{mezard2009information,decelle2011asymptotic,coja2017information} for examples. One can thus expect that also the algorithm of \cite{BIofEpi} is Bayes-optimal on sparse random graphs and that there does not exist an algorithm that finds, on average, a more accurate solution.

In this paper, we investigate whether the BP for inference in spreading models \cite{BIofEpi} provides Bayes-optimal inference on large sparse random graphs. In particular, we re-derive the algorithm, illustrating that it indeed is an instance of a standard BP algorithm applied in a setting where it is expected to be asymptotically optimal. We then investigate in detail its convergence from both random and informed initializations seeking a possible algorithmically hard phase, which we do not find. We study the phase diagram of inference problems, identifying regions of parameters where the optimal estimation error is large and others where it is rather small. 
We also evaluate the so-called Nishimori conditions \cite{nishimori2001statistical, 10.1080/00018732.2016.1211393} that must be satisfied for the Bayes-optimal inference. 
To facilitate the broad usage of these types of algorithms, we also share a basic Python code \cite{bpepi_github} to solve several inference problems that can be easily modified to deal with different spreading models. 

This paper is structured as follows: in Sec.~\ref{sec:BaySet} we introduce the definition and the assumptions on the spreading and observations models; then the Bayesian approach to inference problems is presented. Furthermore, we define the observables and their optimal estimators to measure the performances of the algorithm, the concept of Bayes optimal inference, and the Nishimori conditions. In Sec.~\ref{sec:BPinference} the BP equations are derived and in Sec.~\ref{sec:BOresults} we show the results of the algorithm on some specific inference problems and the corresponding phase diagrams. 

Finally, in Sec.~\ref{sec:Nishi} we focus on some small regions of parameters for which the BP algorithm does not seem Bayes-optimal, in the sense that it does not converge and the Nishimori conditions are no longer respected. We conjecture that this behaviour is due to finite-size effects.

\section{\label{sec:BaySet}Bayesian Inference for local spreading models on networks}

\subsection{Local spreading models on networks}
We consider a network or equivalently a graph $G(V,E)$ with $N=|V|$ nodes. Each node $i$ carries a state variable $x_i^t$ that may change at every time-step $t=0,1,2,\dots,T$. The variable $x_i^t$ can take values from an ordered set of $p+1$ elements. We consider a class of models where at most $p$ changes are possible within the considered time range and each change must be into the consecutive state, according to the ordering of the set. 

For such a class of models (examples below) we represent the time trajectory of a node $i$ by a set of times $\vec{t}_i = \{ t_i^1,\dots,t_i^{p} \}$ defined as the last time steps in which the node is in a certain state (i.e. it will be in a different state the next time step). The entire evolution of the system $\vec{\mathbf{x}}$, of variables $ \{x_i^t\}_{i=1,\dots ,N}^{t=0,\dots, T-1}$, can thus be mapped to the variables $\vec{\mathbf{t}} = \{\vec{t}_i\}_{i=1 ,\dots ,N}$.

In this work, we consider all stochastic spreading processes for which the probability of the evolution can be written in the form  
\begin{equation}\label{eq:CSMprior}
    P(\vec{\mathbf{t}}|\bm{\lambda}) = \frac{1}{Z_{\text{prior}}} \prod_{i=1}^N \Psi_i\left(\vec{t}_i, \{\vec{t}_j\}_{j\in\partial_i}|\bm{\lambda}\right)\,,
\end{equation}
where $Z_{\text{prior}}$ is the normalization, $\bm{\lambda}$ is a set of parameters and $\partial i$ is the set of neighbours of the node $i$ in the graph $G(V,E)$. 
The function $\Psi_i$ can be an arbitrary non-negative function. In this work, we assume the spreading model (\ref{eq:CSMprior}) that governs the process is known, and it depends on a set of parameters $\bm{\lambda}$ that is assumed known. 

\paragraph{Example 1: SIR} A canonical model for epidemic spreading on a network is the SIR model, where each individual node can be in one of the 3 states: susceptible (S), infectious (I), and recovered (R). The spreading process is stochastic and, more specifically, a Markov Chain: at time step $t$, each infectious node $i$ may infect any one of its susceptible contacts $\big\{ j \in \partial_i : x_j^t = S \big\} $ with probability $\lambda_{ij}(t) \in [0,1]$. We call $t_i$ the time of infection of node $i$, i.e. $x_i^{t_i}=S$ while $x_i^{t_i+1}=I$. Then, for each time $t>t_i$, node $i$ can recover (i.e. change to the state $R$, earliest we can have $x_i^{t_i+2}=R$) with probability $\mu_i \in [0,1]$, and again we define the time of recovery $r_i$ as the last time $t$ in which $x_i^{t}=I$. Any given node can only undergo the transitions $S \longrightarrow I \longrightarrow R$. 
We can easily convince ourselves that this epidemic model can be described by the model~(\ref{eq:CSMprior}), by taking $\vec{t}_i = \{t_i, r_i\}$, and as $\bm{\lambda}$ the set of all parameters $\lambda_{ij}(t)$ and $\mu_i$. We also note that the SI model where there is no recovered state is easily obtained as a special case of the SIR with $\mu_i=0$ for every node $i$. 

\paragraph{Example 2: deterministic-SIR}
More complicated models can be considered, for instance, non-Markovian models. As an example, in this work, we consider a slight variant of the SIR model, which we call \textit{deterministic-SIR} (dSIR) or \textit{SIR with deterministic recovery}. The model is no longer a Markov chain and is characterised by a single stochastic transition, i.e. the one between the S and the I states, which stays the same as the SIR model. The second transition, between the $I$ and the $R$ states, is  modelled differently than in the standard SIR model. 
After being infected, the node $i$ remains infectious for a fixed time $\Delta_i$, and then becomes recovered: in this case the infectivity of individual $i$ switches to zero after $\Delta_i$ time steps, and thus there is no stochasticity in the recovery process. In the following, we will consider the case in which $\Delta_i = \Delta\;\forall\,i$. In general, we can consider arbitrarily complex cases in which the infectivity $\lambda_{ij}(t,t-t_i)$ (i.e. the probability of node $i$ infecting node $j$ at time $t$, if $j$ is susceptible and $i$ is infectious) changes both in time and with the time-delay from the time of infection $t_i$.

\paragraph{Example 3: SIS model with up to $\lceil\frac{p}{2}\rceil-1$ reinfections.} 
The SIS model is another very commonly considered model for epidemic spreading, where the infected nodes return to a susceptible state after some time. The SIS model also falls under the framework considered in (\ref{eq:CSMprior}) if we restrict the number of possible re-infections. In particular, if we assume that only $\lceil\frac{p}{2}\rceil-1$ reinfections are possible, we can again formulate the spreading in terms of the $p$-dimensional time trajectories $\vec t_i$ and the probability over them given by (\ref{eq:CSMprior}). Such a model is suitable for infections where immunity is temporary or non-existent, assuming that only a negligible fraction of the population gets reinfected more than $\lceil\frac{p}{2}\rceil-1$  times. E.g. for COVID-19 this seems a reasonable assumption for say $p \approx 20$.

\subsection{The Bayesian inference framework} 

The high-level idea of Bayesian inference for spreading processes is that only some partial observations about the spreading are available on a given network, and the aim is to recover as much information about the spreading as possible. 

We now define what kind of observations we consider. 
We assume site-dependent, factorised observations on the system $\bm{\mathcal{O}} = \{\mathcal{O}_i\}_{i=1,\dots,N}$, that are also independent of the parameters $\bm{\lambda}$ and the other transition times $\vec{\boldsymbol{t}}\,\backslash\vec{t_i}$ when conditioned on the transition times of the node $\vec{t_i}$. Then the likelihood of a set of observations can be written as  $P(\bm{\mathcal{O}}|\vec{\mathbf{t}},\bm{\lambda}) = \prod_{i =1}^NP(\mathcal{O}_i|\vec{t}_{i})$, with $P(\mathcal{O}_i|\vec{t}_{i})$ a known probabilistic law. In this notation, $\mathcal{O}_i$ includes all the partial information we get from the observations on the node $i$, and the case in which the node is not observed corresponds to $\mathcal{O}_i = \varnothing$, for which trivially $P(\mathcal{O}_i = \varnothing | \vec{t}_{i}) = 1 \; \forall\,\vec{t}_{i}$. Instead, if for instance, we observe that a node $i$ has been infected at time $T_{\rm obs}$, we have $P(\mathcal{O}_i|\vec{t}_{i}) = \Ids[t_i=T_{\rm obs}]$, where $\Ids[\cdot]$ is the identity function that is equal to one when the condition in the argument is satisfied and zero otherwise.

In a Bayesian setting, considering $P(\vec{\mathbf{t}}|\bm{\lambda})$ as the prior and $P(\bm{\mathcal{O}}|\vec{\mathbf{t}},\bm{\lambda})$ as the likelihood, we can then recover the configuration through the posterior probability distribution that from the Bayes rule reads 
\begin{align}
\label{eq:post_eq}
   P(\vec{\mathbf{t}}|\bm{\mathcal{O}},\bm{\lambda}) = & \frac{1}{P(\bm{\mathcal{O}}|\bm{\lambda})}P(\vec{\mathbf{t}}|\bm{\lambda})P(\bm{\mathcal{O}}|\vec{\mathbf{t}},\bm{\lambda}) \\
    = & \frac{1}{Z\left(\bm{\mathcal{O}}\right)}\prod_{i=1}^N \Psi_i\left(\vec{t}_i, \{\vec{t}_j\}_{j\in\partial_i}| \bm{\lambda}\right)\prod_{i =1}^NP(\mathcal{O}_i|\vec{t}_{i})\\
    = & \frac{1}{Z\left(\bm{\mathcal{O}}\right)} \prod_{i=1}^N \widetilde{\Psi}_i\left(\vec{t}_i, \{\vec{t}_j\}_{j\in\partial_i},\mathcal{O}_i| \bm{\lambda}\right)\,,
\end{align}
where we defined the normalization constant $Z\left(\bm{\mathcal{O}}\right) \equiv Z_{\rm prior}P\left(\bm{\mathcal{O}}|\bm{\lambda}\right) = Z_{\rm prior}\sum_{\vec{\mathbf{t}}} \prod_{i=1}^N \widetilde{\Psi}_i\left(\vec{t}_i, \{\vec{t}_j\}_{j\in\partial_i},\mathcal{O}_i| \bm{\lambda}\right)$ and 
$\widetilde{\Psi}_i\left(\vec{t}_i, \{\vec{t}_j\}_{j\in\partial_i},\mathcal{O}_i| \bm{\lambda}\right) \equiv \Psi_i\left(\vec{t}_i, \{\vec{t}_j\}_{j\in\partial_i}| \bm{\lambda}\right)P\left(\mathcal{O}_i|\vec{t}_i\right)$. 

Given the observations, there are different properties of the spreading that we may want to infer. For instance, one aim is to identify the sources of the infection (the patient(s) zero problem). This reduces to estimating the marginals over the individuals (variables) to be infectious at time step zero. Another possible goal is to assess the epidemic risk of being infectious at time $t$, for this, we need to compute the probability for each variable to be infectious at time $t$. A broad range of other possible goals reduces to computing marginals of the posterior, or averages of quantities over the posterior probability distribution. 

The number of configurations of the systems grows exponentially with the system size, making the exhaustive computation of marginals and averages  impossible for a size larger than a few dozen of nodes.  In this work, we aim to show how the Belief Propagation (BP) equations are used to estimate efficiently the marginals of this posterior probability distribution in a range of problems where the network of interactions among the variables is random and sparse.

\subsubsection{The optimal overlap} 
Let us define the overlap $\text{O}(\mathbf{x}, \mathbf{y})$ between two vectors $\mathbf{x}$ and $\mathbf{y}$, of discrete entries and  length $N$, as the fraction of agreeing elements between the two:
\begin{equation}\label{eq:ov}
    \text{O}(\mathbf{x}, \mathbf{y}) = \frac{1}{N} \sum_{i=1}^{N} \delta_{x_i,y_i}\,.
\end{equation}
In particular we are interested in the quantity $\text{O}(\hat{\mathbf{x}}^t, \mathbf{x}^{*,t})$, where $\mathbf{x}^{*,t}$ is the ground-truth of the state of the system at time~$t$, and $\hat{\mathbf{x}}^t$ is a generic estimator of it. In general, we do not know the ground truth; assuming that it is distributed accordingly to the posterior, the best thing we can do (in a Bayesian framework) is to estimate the so-called Mean Overlap:
\begin{equation}\label{eq:MO}
    \text{MO}(\hat{\mathbf{x}}^t) = \frac{1}{N} \sum_{\mathbf{x}^t} P(\mathbf{x}^t | \bm{\Ocl}) \sum_{i=1}^{N} \delta_{x_i^t,\hat{x}_i^t}\,,
\end{equation}
where $P(\mathbf{x}^t | \bm{\Ocl})$ can be computed from $P(\vec{\mathbf{t}}|\bm{\mathcal{O}}x)$. Maximizing this quantity over the estimator $\hat{\mathbf{x}}$ leads to the \textit{maximum mean overlap estimator}
\begin{equation}
    \hat{x}_i^{t,\text{MMO}} = \argmax_{x_i^t} P_i(x_i^t| \bm{\Ocl})\,
\end{equation}
where $P_i(x_i^t| \bm{\Ocl}) = \sum_{\mathbf{x}^t\backslash x_i^t}P(\mathbf{x}^t | \bm{\Ocl})$ is the marginal probability for the node $i$ to be in a given state at time $t$. In the following, we will refer to the performances of this estimator simply as $\text{O}_t \equiv \text{O}(\hat{\mathbf{x}}^{t,\text{MMO}}, \mathbf{x}^{*,t})$ and $\text{MO}_t \equiv \text{MO}(\hat{\mathbf{x}}^{t,\text{MMO}})$, both implicitly depending on the marginal probability distributions.

Starting from these definitions of overlap and mean overlap, we now define the performance parameters which we are going to use in our numerical studies. The idea is to rescale (\ref{eq:ov}) and (\ref{eq:MO}) by comparing the performance of the MMO estimator to the one obtained by running BP without observations, i.e. on the prior, which in the following we will call \textit{random estimator} and can be written as 
\begin{equation}
    \hat{x}_i^{\text{RND},t} = \argmax_{x_i^t} P_i(x_i^t)\,.
\end{equation}
where $P_i(x_i^t) = \sum_{\mathbf{x}^t\backslash x_i^t}P(\mathbf{x}^t )$ is the marginal of the prior probability. We  write the resulting estimate of the state of the system at time $t$ as $\hat{\mathbf{x}}^{\text{RND},t}$ and we then define the \textit{rescaled overlap} as
\begin{equation}\label{eq:ResOv}
    \widetilde{\text{O}}_t  = \frac{\text{O}_t - \text{O}(\hat{\mathbf{x}}^{\text{RND},t}, \mathbf{x}^{*,t})}{1 - \text{O}(\hat{\mathbf{x}}^{\text{RND},t}, \mathbf{x}^{*,t})}
\end{equation}
and the \textit{rescaled mean overlap} as
\begin{equation}\label{eq:ResMOv}
    \widetilde{\text{MO}}_t = \frac{\text{MO}_t - \text{MO}(\hat{\mathbf{x}}^{\text{RND},t})}{1 - \text{MO}(\hat{\mathbf{x}}^{\text{RND},t})}\,.
\end{equation}
Notice how these performance parameters are defined in such a way that they are negative or zero if and only if BP performs worse or equal to the random estimator, respectively, and are in $(0,1]$ in any other case, with the value one indicating a perfect resolution of the problem.

\subsubsection{The optimal mean squared error}
Another property we want to study is how well our algorithm can estimate the entire trajectory of each node. We will focus on models that feature a single transition time, such as the SI and dSIR models. However, one can extend our analysis to cover models with multiple transition times, such as the SIR and SIS models, by generalizing the following definitions. We consider the squared error between the ground-truth vector $\mathbf{t}^*$ and an estimator $\hat{\mathbf{t}},$ defining
\begin{equation}\label{eq:se}
    \text{SE}(\hat{\mathbf{t}}, \mathbf{t}^*) = \frac{1}{N}\sum_{i=1}^{N} (\hat{t}_i - t_i^*)^2\,.
\end{equation}
As before, since we usually do not know the ground truth, we can assume that $\mathbf{t}^*$ is distributed according to the posterior probability distribution and then define the \textit{mean squared error} (averaged on the posterior) as
\begin{equation}\label{eq:MSE}
    \text{MSE}(\hat{\mathbf{t}}) = \frac{1}{N} \sum_{\mathbf{t}} P(\mathbf{t} | \bm{\Ocl})\sum_{i=1}^N (\hat{t}_i - t_i)^2\,.
\end{equation}
In this case, minimizing Eq.~(\ref{eq:MSE}) with respect to the estimator, we obtain the \textit{minimum mean squared error estimator}
\begin{equation}
    \hat{t}_i^{\text{MMSE}} = \sum_{\mathbf{t}} P(\mathbf{t} | \bm{\Ocl}) t_i = \sum_{t_i} P_i(t_i| \bm{\Ocl}) t_i\,,
\end{equation}
where $P_i(t_i|\bm{\Ocl}) = \sum_{\mathbf{t}\backslash t_i}P(\mathbf{t} | \bm{\Ocl})$ is the marginal probability for the node $i$ to have a certain time of transition $t_i$. In the following, we will refer to the performances of this estimator simply as $\text{SE} = \text{SE}(\hat{\mathbf{t}}^{\text{MMSE}}, \mathbf{t}^*)$ and $\text{MSE} = \text{MSE}(\hat{\mathbf{t}}^{\text{MMSE}})$, both implicitly depending on the marginal probability distributions $P_i(t_i|\bm{\Ocl})$.

We can define the random estimator for the times of transition, as we did for the overlap where we averaged over the prior probability distribution, as
\begin{equation}
    \hat{t}_i^{\text{RND}} = \sum_{t_i} P_i(t_i) t_i\,.
\end{equation}
where $ P_i(t_i) = \sum_{\mathbf{t}\backslash t_i}P(\mathbf{t})$ is the marginal of the prior probability. 
Then, our performance parameters will be, respectively
\begin{equation}\label{eq:RSE}
    R_{\text{SE}}= \frac{\text{SE}(\hat{\mathbf{t}}^{\text{RND}},\mathbf{t}^*) - \text{SE}}{ \text{SE}(\hat{\mathbf{t}}^{\text{RND}},\mathbf{t}^*)}
\end{equation}
and
\begin{equation}\label{eq:RMSE}
    R_{\text{MSE}}= \frac{\text{MSE}(\hat{\mathbf{t}}^{\text{RND}}) - \text{MSE}}{ \text{MSE}(\hat{\mathbf{t}}^{\text{RND}})}\,.
\end{equation}
Notice how, as for the rescaled overlaps, these performance parameters are defined in such a way that they are negative or zero if and only if BP performs worse or equal to the random estimator, respectively, and are in $(0,1]$ in any other case, with the value one indicating a perfect resolution of the problem.

\subsection{Nishimori conditions}
In the Bayes-optimal setting, i.e., when the prior probability distribution and the probability function used to generate the observations are known, it is possible to derive consistency conditions on expectation over the posterior that are called the \textit{Nishimori conditions} in statistical physics of disordered systems \cite{nishimori2001statistical,10.1080/00018732.2016.1211393}. 
In words, the Nishimori conditions state that, under averages over the posterior, we cannot distinguish the ground truth we aim to infer and a random sample from the posterior. 
These conditions imply certain properties of the optimal estimators that we make explicit below. As a consequence, when we consider an algorithm that approximates the optimal estimators, the Nishimori conditions can serve as a necessary condition for the approximation to be close to the true optimal estimators. This is how we will use these conditions in the present work. 

To state the Nishimori conditions, let us consider having a ground-truth configuration $\mathbf{x}^*$ generated from the prior distribution $P_{\text{g.t.}}(\mathbf{x}^*)$ and a measurement process leading to observations $\bm{\mathcal{O}}$, generated through the likelihood function $P_{\text{g.t.}}(\bm{\mathcal{O}}|\mathbf{x}^*)$. Let $\mathbf{x}$ denote a sample from the posterior $P(\mathbf{x}|\bm{\mathcal{O}})$. Then, given an observable $f(\mathbf{x})$, we can compute the two expectations:
\begin{subequations}
\begin{align}
    \mathbb{E}_{\mathbf{x}^*}[f(\mathbf{x}^*)] = \sum_{\mathbf{x}^*} f(\mathbf{x}^*) P_{\text{g.t.}}(\mathbf{x}^*) = \sum_{\mathbf{x}^*,\,\bm{\mathcal{O}} } f(\mathbf{x}^*) P_{\text{g.t.}}(\mathbf{x}^*)
    P_{\text{g.t.}}(\bm{\mathcal{O}}|\mathbf{x}^*) \label{eq:Nish1}\\
    \mathbb{E}_{\bm{\mathcal{O}}}\mathbb{E}_{\mathbf{x}|\bm{\mathcal{O}}}[f(\mathbf{x})] = \sum_{\mathbf{x},\,\bm{\mathcal{O}} } f(\mathbf{x}) P(\bm{\mathcal{O}})P(\mathbf{x}|\bm{\mathcal{O}}) = \sum_{\mathbf{x},\,\bm{\mathcal{O}} } f(\mathbf{x}) P(\mathbf{x})P(\bm{\mathcal{O}}|\mathbf{x})\label{eq:Nish2}\,,
\end{align}
\end{subequations}
where in the first equation we used the fact that $\sum_{\bm{\mathcal{O}} } P_{\text{g.t.}}(\bm{\mathcal{O}}|\mathbf{x}^*) = 1$ and in the second we applied Bayes theorem. Now we can notice that $\mathbf{x}$ and $\mathbf{x}^*$ are two dummy variables, such that when $P(\mathbf{x}) = P_{\text{g.t.}}(\mathbf{x})$ and $P(\bm{\mathcal{O}}|\mathbf{x}) = P_{\text{g.t.}}(\bm{\mathcal{O}}|\mathbf{x})$ the two expressions~(\ref{eq:Nish1}) and~(\ref{eq:Nish2}) coincide, and we say that the Nishimori conditions are satisfied. Therefore, checking that $\mathbb{E}[f(\mathbf{x}^*)] = \mathbb{E}[f(\mathbf{x})]$ can give us solid evidence that we are in the \textit{Bayes-optimal} case, and thus that the solution the algorithm finds is the best possible in a Bayesian framework.

For the overlaps defined above, the Nishimori conditions lead to checking that the two following quantities coincide:
\begin{subequations}
\label{eq:NishiOV}
\begin{align}
    \E[\text{O}_t] &= \frac{1}{N} \sum_{i=1}^{N} \E\big[\Ids[\argmax_{x_i^t} P_i(x_i^t|\bm{\Ocl}) = x_i^{*,t}]\big]\,, \\
\begin{split}
    \E\big[\text{MO}_t \big] &= \frac{1}{N} \sum_{i=1}^{N} 
    \E\big[\sum_{\widetilde{x}^{t}_i} P_i(\widetilde{x}^{t}_i|\bm{\Ocl}) \Ids[\argmax_{x_i^t} P_i(x_i^t|\bm{\Ocl}) = \widetilde{x}^{t}_i]\big] \\
    &= \frac{1}{N}\sum_{i=1}^{N} \E\big[ \max_{x_i^t}P_i(x_i^t|\bm{\Ocl})\big]\,.
\end{split}
\end{align}
\end{subequations}
Analogously, for the squared errors:
\begin{subequations}
\begin{align}
    \E\big[ \text{SE} \big] &= \frac{1}{N}\sum_{i=1}^{N} \E\left[\left(\sum_{t_i}P_i(t_i|\bm{\Ocl})t_i - t_i^*\right)^2\right]\,, \\
    \E \big[ \text{MSE} \big] &= \frac{1}{N}\sum_{i=1}^{N} \E\left[\sum_{t'_i}P_i(t'_i|\bm{\Ocl})\left(\sum_{t_i}P_i(t_i|\bm{\Ocl})t_i - t'_i\right)^2\right]\,.
\end{align}
\end{subequations}
The above equalities tell us that, in the Bayes-optimal case, on average, the mean of the observable computed from the posterior must be equal to the actual observable calculated on the ground truth. In the following, we will show some cases where the BP algorithm, described in Section~\ref{sec:BPinference}, respects them, and we are confident that the solutions found are the best possible on average. Then in Section \ref{sec:Nishi}, we will show other cases where the BP equations struggle to converge, the Nishimori conditions are no longer satisfied, and other algorithms should be used if one aims to reach Bayes optimality.

\section{\label{sec:BPinference}Inference with Belief Propagation}

The belief propagation (BP) equations allow computing marginals exactly when the factor graph associated with a probability distribution is acyclic~\cite{yedidia2003understanding}. 
In this paper, we consider large random graphs that are locally tree-like, i.e. the typical length of loops grows as $\log N$, with $N$ the number of nodes. This is the case of Erdős–Rényi or random regular graphs (RRG) with an average degree of order one~\cite{bollobas1998random}. 

In random locally tree-like factor graphs, the BP algorithm is expected to estimate the marginals asymptotically exactly, i.e. with an error vanishing in the limit of $N \rightarrow \infty$, as long as the system is in the so-called replica symmetric (RS) phase. 
As long as we are dealing with inference problems in a Bayes optimal setting, such as we are considering here, the equilibrium properties are in the replica symmetric phase. This follows from the Nishimori conditions. We thus expect the BP equations to give the correct marginals as long as their fixed point satisfies the Nishimori conditions and in the absence of a first-order phase transition that may prevent the BP algorithm from reaching the corresponding fixed point~\cite{10.1080/00018732.2016.1211393}.
The presence of a first-order phase transition can be verified by running the BP algorithm both from random and from informed initializations and checking whether they converge to the same fixed point. 

\begin{figure}[t!]
    \centering
    \includegraphics[width=0.8\linewidth]{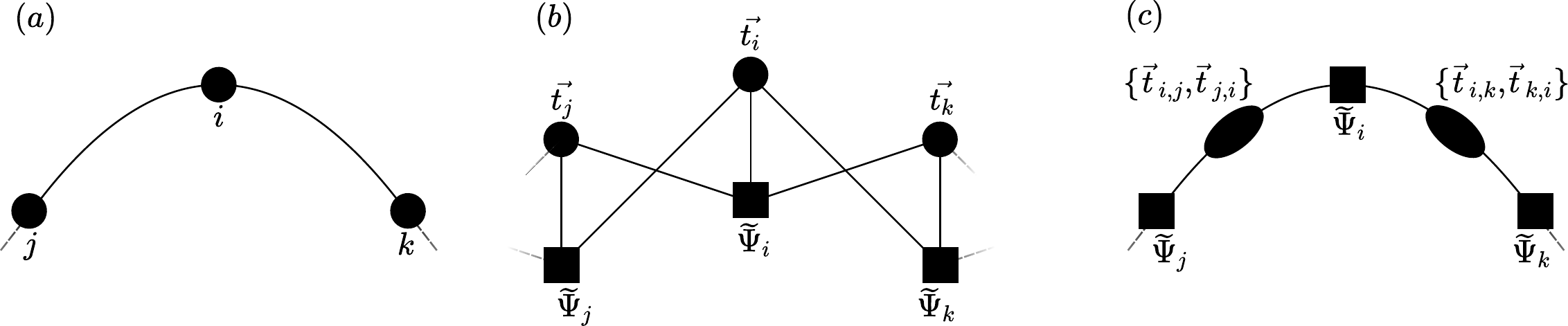}
    \caption{\label{fig:FacGraph}\textbf{Factor graph representation of the posterior~(\ref{eq:post_eq}).} In (a) we show a possible path in a contact network, and in (b) its respective naive factor graph representation. In (c) we show the representation obtained by grouping the variables as explained in the main text. }
\end{figure}

In our setting, however, the naive factor graph representation of the posterior probability distribution~(\ref{eq:post_eq}), where the nodes are the time trajectories of each variable and the factors are the set of $\{\widetilde{\Psi}_i\}$, presents small loops even when the interaction graph among variables is acyclic. The reason is that the factor $\widetilde{\Psi}_i$ couples $i$ together with all its neighbours $k \in \partial i$, see Figure~\ref{fig:FacGraph}, panel (b).
Instead, we use a different factor graph representation, introducing additional variables. We replicate the variables $\vec{t}_i$ on each edge of the interaction graph, calling the replicas obtained $\vec{t}_{i,j}$ with $j \in \partial i$, and we add local constraints imposing their equality. Calling $\vec{\mathbf{t}}_{\bm{\partial}} = \{ \vec{t}_{i,j} \}_{i=1,\dots,N}^{j \in \partial i}$, the new representation of the posterior probability distribution is: 
\begin{equation}
    P(\vec{\mathbf{t}},\vec{\mathbf{t}}_{\bm{\partial}}|\bm{\mathcal{O}},\bm{\lambda}) =  \frac{1}{Z\left(\bm{\mathcal{O}}\right)} \prod_{i=1}^N \widetilde{\Psi}_i\left(\vec{t}_i, \{\vec{t}_{j,i}, \vec{t}_{i,j}\}_{j\in\partial_i}| \bm{\lambda}\right) \prod_{j\in \partial i} \Ids  \left[ \vec{t}_{i,j} = \vec{t}_i \right]\,,
\end{equation}
and the respective factor graph is represented in Figure~\ref{fig:FacGraph}, panel (c). The BP equations associated with this posterior distribution are:
\begin{equation}\label{eq:BP_gen_dupl}
    m_{i\rightarrow j}(\vec{t}_{i,j},\vec{t}_{j,i}) =
    \frac{1}{Z_{i\rightarrow j}}
    \sum_{\substack{\vec{t}_i, \\ \{ \vec{t}_{i,k}, \vec{t}_{k,i} \}_{k\in \partial_i\backslash j}}}
    \widetilde{\Psi}_i\left(\vec{t}_i, \{\vec{t}_{k,i}, \vec{t}_{i,k}\}_{k\in\partial_i}| \bm{\lambda}\right)
    \Ids  \left[ \vec{t}_{i,j} = \vec{t}_i \right]
 \prod_{k\in\partial_i\backslash j}
  \Ids  \left[ \vec{t}_{i,k} = \vec{t}_i \right]
  m_{k\rightarrow i} ( \vec{t}_{i,k},\vec{t}_{k,i})\,,
\end{equation}
where the message $m_{i\rightarrow j}$ goes from the factor node $\widetilde{\Psi}_i$ to $\widetilde{\Psi}_j$ and 
$$
Z_{i\rightarrow j} =
\sum_{\substack{\vec{t}_i, \\ \{ \vec{t}_{i,k}, \vec{t}_{k,i} \}_{k\in \partial_i}}}
    \widetilde{\Psi}_i\left(\vec{t}_i, \{\vec{t}_{k,i}, \vec{t}_{i,k}\}_{k\in\partial_i}| \bm{\lambda}\right)
    \Ids  \left[ \vec{t}_{i,j} = \vec{t}_i \right]
 \prod_{k\in\partial_i\backslash j}
  \Ids  \left[ \vec{t}_{i,k} = \vec{t}_i \right]
  m_{k\rightarrow i} ( \vec{t}_{i,k},\vec{t}_{k,i})
$$ 
is the normalization constant.
By imposing the indicator functions and renaming the dummy variables we obtain the final form,
\begin{equation}\label{eq:BP_gen}
    m_{i\rightarrow j}(\vec{t}_i,\vec{t}_j) =
    \frac{1}{Z_{i\rightarrow j}}
    \sum_{ \{\vec{t}_k \}_{k\in \partial_i\backslash j}}
    \widetilde{\Psi}_i\left(\vec{t}_i, \{\vec{t}_k\}_{k\in\partial_i}| \bm{\lambda}\right)
 \prod_{k\in\partial_i\backslash j}
  m_{k\rightarrow i} ( \vec{t}_k,\vec{t}_i)\,.
\end{equation}
The messages are usually initialized randomly or uniformly and then updated according to Eq.~(\ref{eq:BP_gen}) until convergence if they converge, or otherwise when they meet some stopping criterion. In this paper, we will also consider an informed initialization corresponding to the ground truth (that is of course not available in practice) to check for first-order phase transitions appearing in the problem. 
The marginals of each variable are then computed \cite{yedidia2003understanding} as
\begin{equation}\label{eq:BPmarg}
    b_i\left(\vec{t}_i\right) = \frac{1}{Z_i}
    \sum_{ \{\vec{t}_k \}_{k\in \partial_i}}
    \widetilde{\Psi}_i\left(\vec{t}_i, \{\vec{t}_k\}_{k\in\partial_i}| \bm{\lambda}\right)
 \prod_{k\in\partial_i}
  m_{k\rightarrow i} ( \vec{t}_k,\vec{t}_i)\,,
\end{equation}
and from the fixed point one can also compute the log-partition function as 
\begin{equation}
    \log Z = \sum_i \log Z_i - \sum_{(i,j)} \log Z_{(i,j)}\,,
\end{equation}
where
\begin{subequations}
\begin{align}
    Z_i &= \sum_{\vec{t}_i} \sum_{ \{\vec{t}_k \}_{k\in \partial_i}}
    \widetilde{\Psi}_i\left(\vec{t}_i, \{\vec{t}_k\}_{k\in\partial_i}| \bm{\lambda}\right)
 \prod_{k\in\partial_i}
  m_{k\rightarrow i} ( \vec{t}_k,\vec{t}_i)\,, \\
  Z_{(i,j)} &= \sum_{\vec{t}_i}\sum_{\vec{t}_j} m_{i\rightarrow j} ( \vec{t}_i,\vec{t}_j) m_{j\rightarrow i} ( \vec{t}_j,\vec{t}_i)\,.
\end{align}
\end{subequations}
The computational complexity of the brute-force calculation of the marginals is $\mathcal{O}(T^{Np})$, and thus scales exponentially with the system size and the number of possible transitions. Using the BP equations, one update of all the messages reduces to $\mathcal{O}(E\, T^{(d_{max}-1)p})$ where $E$ is the number of edges in the interaction graph considered, and $d_{max}$ is the maximum degree of the nodes in the interaction graph. Now the computation is no longer exponential in the number of variables, but still exponential in the degree of the graph. Depending on the particular spreading model, more efficient and smaller representations of the time trajectories can be adopted to avoid the exponential dependence on the degree. In appendix~\ref{App:BPEPI} the BP equations are explicitly written for the deterministic-SIR model, where this simplification holds.

\section{Bayes Optimal Phase Diagrams}
\label{sec:BOresults}

This section presents the results of applying the BP algorithm, described in Section~\ref{sec:BPinference}, to the dSIR model on epidemic inference problems. We use the definition of Nishimori identities, as provided in Section~\ref{sec:BaySet}, to investigate Bayes-optimality for various observables. The setting of the inference problems is:
\begin{description}
    \item[Network] For simplicity, in the following, we will consider a single ensemble of graphs: the random regular graphs of degree $d=3$, not varying in time. We consider a homogeneous probability of infection $\lambda$ for each contact, equal for each edge of the interaction graph, and constant in time.
    
    \item[Sources] The sources are chosen uniformly at random, i.e. each individual could be infectious at time $t=0$ with probability $\delta$. Thus, the $\delta$ parameter controls our simulations' average fraction of sources.
    \item[Observations] we will focus on two different types of observations, both typically used in the epidemic inference problems setting:
    \begin{description}
        \item[Sensors:] A fraction $\rho$ of individuals, called sensors, are chosen uniformly at random. These nodes reveal their entire trajectory in time, which for the dSIR model means just their time of infection $t_i$. Such observations were considered for instance in \cite{10.1103/physrevlett.109.068702,Spi19}. 
        \item[Snapshot:] The state of all variables at a specific time $T_{\rm obs}$ is known. We get information on whether the node is still susceptible (S) or not (either I or R). Such observations were considered for instance in \cite{InfOri,altarelli2014patient}.
    \end{description}
\end{description}

We will consider two observables, the overlap at time 0, $O_{t=0}$, and the squared error, SE. For the first one, in the numerical results, we compute the two following quantities 
\begin{subequations}
\begin{align}\label{eq:Nishi_O0_BP}
    \overline{\text{O}_{t=0}} &=  \frac{1}{N_{\text{sim}}} \sum_{s=1}^{N_{\text{sim}}} \frac{1}{N} \sum_{i=1}^{N} \Ids\left[\argmax_{x} b_{i,s}^{t=0}(x) = x_{i,s}^{*,t=0}\right]\,, \\
\begin{split}\label{eq:Nishi_MO0_BP}
    \overline{\text{MO}_{t=0}} &= \frac{1}{N_{\text{sim}}} \sum_{s=1}^{N_{\text{sim}}} \frac{1}{N}\sum_{i=1}^{N} \max_{x}b_{i,s}^{t=0}(x)\,.
\end{split}
\end{align}
\end{subequations}
where we wrote the sample mean explicitly introducing the simulation index $s$ and calling $N_{\text{sim}}$ the total number of simulations. The $b^t_{i}(x)$ is the BP estimation of $P_i(x_i^t|\bm{\mathcal{O}})$ that can be easily computed from the $b_i(\vec{t}_i)$.
In the case of the squared error, restricting ourselves to the dSIR model, we need to consider just a single transition time $t_i$ for each node, and the marginals $b_i$ are just one-dimensional vectors.

Analogously to the previous case, we compute:
\begin{subequations}
\begin{align}
    \overline{\text{SE}} &= \frac{1}{N_{\text{sim}}} \sum_{s=1}^{N_{\text{sim}}} \frac{1}{N}\sum_{i=1}^{N} \left(\sum_{t}b_{i,s}(t)t - t_{i,s}^*\right)^2\,, \\
    \overline{\text{MSE}} &= \frac{1}{N_{\text{sim}}} \sum_{s=1}^{N_{\text{sim}}}\frac{1}{N}\sum_{i=1}^{N} \sum_{t'}b_{i,s}(t')\left(\sum_{t}b_{i,s}(t)t - t'\right)^2\,.
\end{align}
\end{subequations}
Then, suppose we are in the Bayes-optimal case. In that case, we must observe that (for $N_{\text{sim}}$ sufficiently large) these two couples of average values are equal up to their statistical error, $(\overline{\text{O}_{t=0}}=\overline{\text{MO}_{t=0}})$ and $(\overline{\text{SE}} = \overline{\text{MSE}})$. 
In practice, in the following, we show the results for the set of rescaled  observables $(\widetilde{\text{O}}_{t=0}, \widetilde{\text{MO}}_{t=0}, R_{SE},  R_{\rm MSE})$ defined in Sec.~\ref{sec:BaySet}.

Two different initializations of the BP messages are used:
\begin{description}
    \item[rnd] we let the BP algorithm converge at a fixed point in the case of no observations. Then we use the resulting values of the messages as the initial point,
    \item[inf] we let the BP algorithm converge at a fixed point where the times of infection of all individuals are observed. Then we use the resulting values of the messages as the initial point.
\end{description}

If the algorithm converges to the same fixed point for both of the two initializations, it indicates an absence of a first-order phase transition and asymptotic optimality of the obtained results, as in e.g.~\cite{decelle2011asymptotic}. Furthermore, checking the validity of the Nishimori conditions that hold on average in the large size limit provides another evidence that the BP algorithm reached the Bayes-optimal marginals. In the following, we will analyse phase diagrams of various inference settings and for each point, we will check the two initializations and the validity of the Nishimori conditions. 

\begin{figure}[t!]
    \centering
    \includegraphics[width=\textwidth]{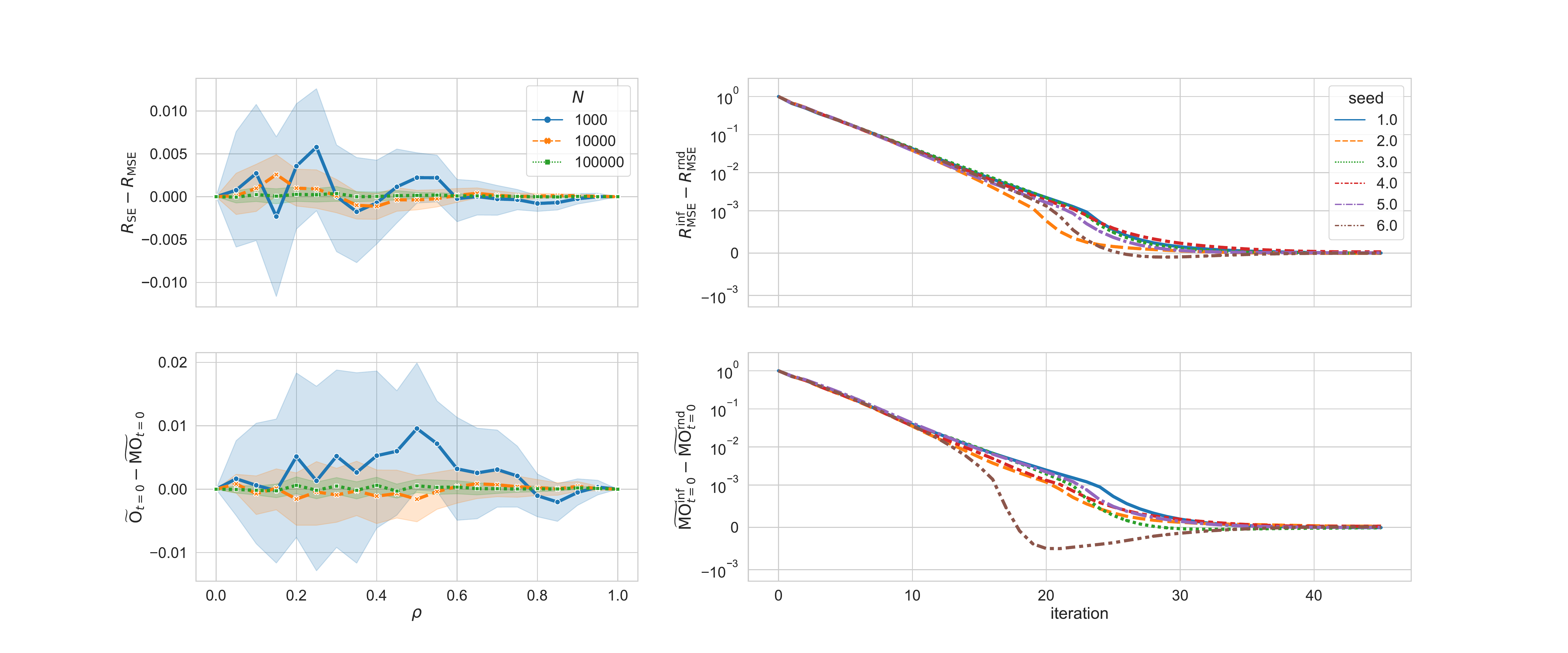}
    \caption{\label{fig:SumNishiSec5}\textbf{Nishimori conditions and initializations difference.} The setting is the following: we consider SI epidemics with sensors on an ensemble of 3-RRGs with $\lambda=0.8$ and $\delta=0.1$. In the left panels, we furthermore vary $N$ in $[10^3,10^5]$ and $\rho$ in $[0,1]$, plotting the Nishimori differences $R_{SE}- R_{\rm MSE}$ in the upper panel and $\widetilde{\text{O}}_{t=0} - \widetilde{\text{MO}}_{t=0}$ in the lower panel, both computed with a \textit{rnd} initialization. The averages are done over $50$ instances. Instead, in the right panels we fix $N=10^4$ and $\rho=0.2$ and we track BP through its iterations until convergence for six randomly generated instances of the problem, for each starting both with \textit{rnd} and \textit{inf} initializations. Using as reference the observables defined in (\ref{eq:RSE}) and (\ref{eq:ResOv}), we plot the difference between the two initializations, and we show that they lead, for each instance, to the same result. The scale of the y-axis of these plots is set as \textit{symlog}, meaning it is logarithmic outside the interval $[-10^{-3},10^{-3}]$ around zero, in which it is linear. The damping parameter $\eta$ was fixed to $0.4$ for all iterations (see Appendix \ref{sec:damp} for details).}
\end{figure}

In Fig.~\ref{fig:SumNishiSec5} we illustrate such a check that is later done for each point in the phase diagrams:
\begin{itemize}
    \item In the left panels, we probe the Nishimori conditions through the rescaled observables defined in Sec.~\ref{sec:BaySet}, namely computing $R_{\rm SE}- R_{\rm MSE}$ and $\widetilde{\text{O}}_t - \widetilde{\text{MO}}_t$ on the single instances and then averaging the values found. By doing this for different sizes $N$, we see how, as expected, increasing the size decreases the variance of the differences and concentrates the single-instance values to zero.
    \item In the right panels, we track the BP algorithm through its iterations for some random instances of the problem, initializing first the messages randomly and then repeating the experiment initializing the messages informatively (see details above). We can see how in each instance showed the two initializations start converging exponentially to the same value after just a dozen of iterations until the difference becomes of the order of the tolerance imposed to check the convergence of BP.
\end{itemize}

\subsection{Phase diagrams for observations via sensors}
\subsubsection{Finding the sources}
\begin{figure}[t!]
    \centering
    \includegraphics[width=0.85\textwidth]{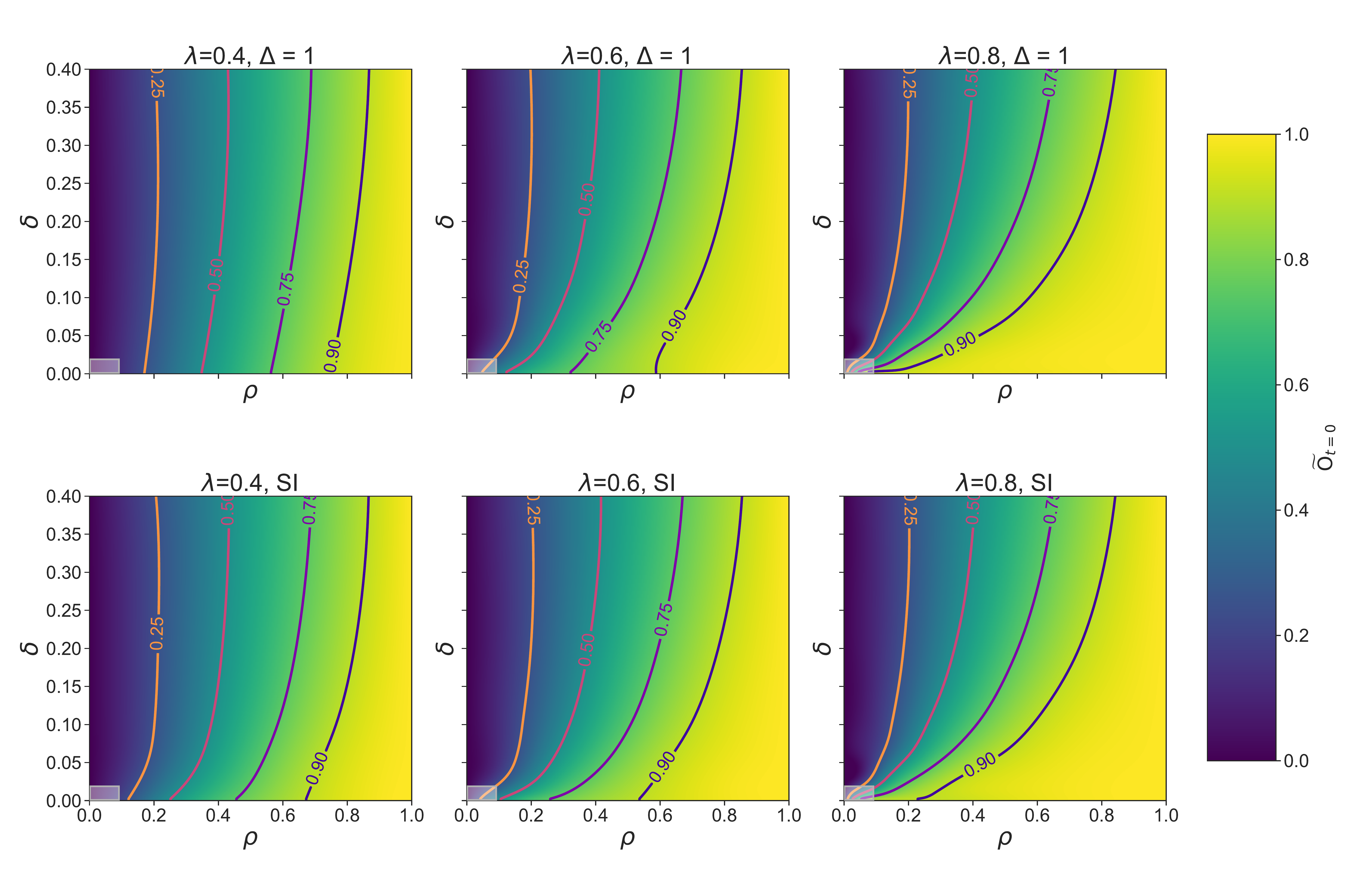}
    \caption{\label{fig:PhDiag_Ot}\textbf{Phase diagrams for source-finding via sensors.} We show the rescaled overlap defined in Eq.~(\ref{eq:ResOv}) for 3-RRGs with $N=10^4$, computed at time $t=0$. We juxtapose the SI model to the dSIR model with $\Delta=1$, and for both, we compare three different values of $\lambda$, and we show how the performance parameter varies with the fraction of sensors $\rho\in[0,1]$ and the fraction of sources $\delta\in[2.5\cdot 10^{-3},0.4]$. The plots show the results for the case of random initialization since they are practically indistinguishable from the informed ones. For all the points outside the grey rectangles, the Nishimori conditions were verified, and thus also $\widetilde{\text{MO}}_{t=0}$ has the same behaviour. Each contour plot was obtained by fitting the value of the rescaled overlap over a $21\times24$ grid of data, averaging 25 simulations for each point on the grid. In the regions at small values of $\delta$ and $\rho$ enclosed in the grey rectangles, we observe that the BP equations no longer converge, the Nishimori identities do not hold, and the results are no longer Bayes Optimal. More insights about these regions are covered in Sec.~\ref{sec:Nishi}. }
\end{figure}

In this section, the performance in finding the sources of infection on random regular graphs is analysed. The results of rescaled overlap $\widetilde{\text{O}}_{t=0}$ computed with the BP algorithm are shown in Fig.~\ref{fig:PhDiag_Ot}. 
The plots show the phase diagram of the problem for different systems settings. We consider both an SI model and a dSIR model with delay $\Delta=1$, and, as stopping criteria of the simulation, when all the individuals are infected and when there are no more infectious individuals, respectively. Fixing $\lambda=0.4,0.6,0.8$, we vary $\delta$ and $\rho$ inside an interesting interval. 

In general, we can see that the system presents a transition between a phase (at low $\delta$ and high $\rho$) where the inference is successful, and we can find (most of) the sources, to another one (at high $\delta$ and low $\rho$) where the inference is unfeasible since we reach the performance of the random estimator. Varying $\lambda$ and $\Delta$ we observe several interesting phenomena:
\begin{itemize}
    \item Decreasing the probability of transmission $\lambda$, starting from $\lambda = 0.8$,  we see that the zone we infer as the random estimator remains more or less the same. At the same time, the zone of perfect inference shirks. In particular, it moves more and more to the right, and we need more and more observations to achieve the same performance. Furthermore, it is interesting to notice that this effect is visible, especially at low values of the parameter $\delta$, where the level lines (in which the rescaled overlap has a fixed value) become more and more equally distributed, which in other terms tells us that the transition becomes less and less sharp. Instead, this effect disappears when the fraction of sources increases and the performance parameter becomes less dependent on the value of $\lambda$.
    \item The inference on the sources depends just slightly on the parameter $\Delta$. 
    We notice that the two cases (on the lower and upper panels of Fig.~\ref{fig:PhDiag_Ot} respectively) are very similar, especially for high values of the parameter $\delta$.
    \item There is a small region of parameters, at small values of $\rho$ and $\delta$, where the BP equations struggle to converge, the Nishimori identities are no longer respected, and we observe a dependence of the results on the initialization of the BP messages. In this region, we can claim that the BP equations do not solve the problems in a Bayes Optimal way. In Sec.~\ref{sec:Nishi}, a detailed analysis of this and other regions where the BP algorithm poorly perform is shown.  
\end{itemize}

\subsubsection{Inference on the times of infection}
\begin{figure}
    \centering
    \includegraphics[width=0.85\textwidth]{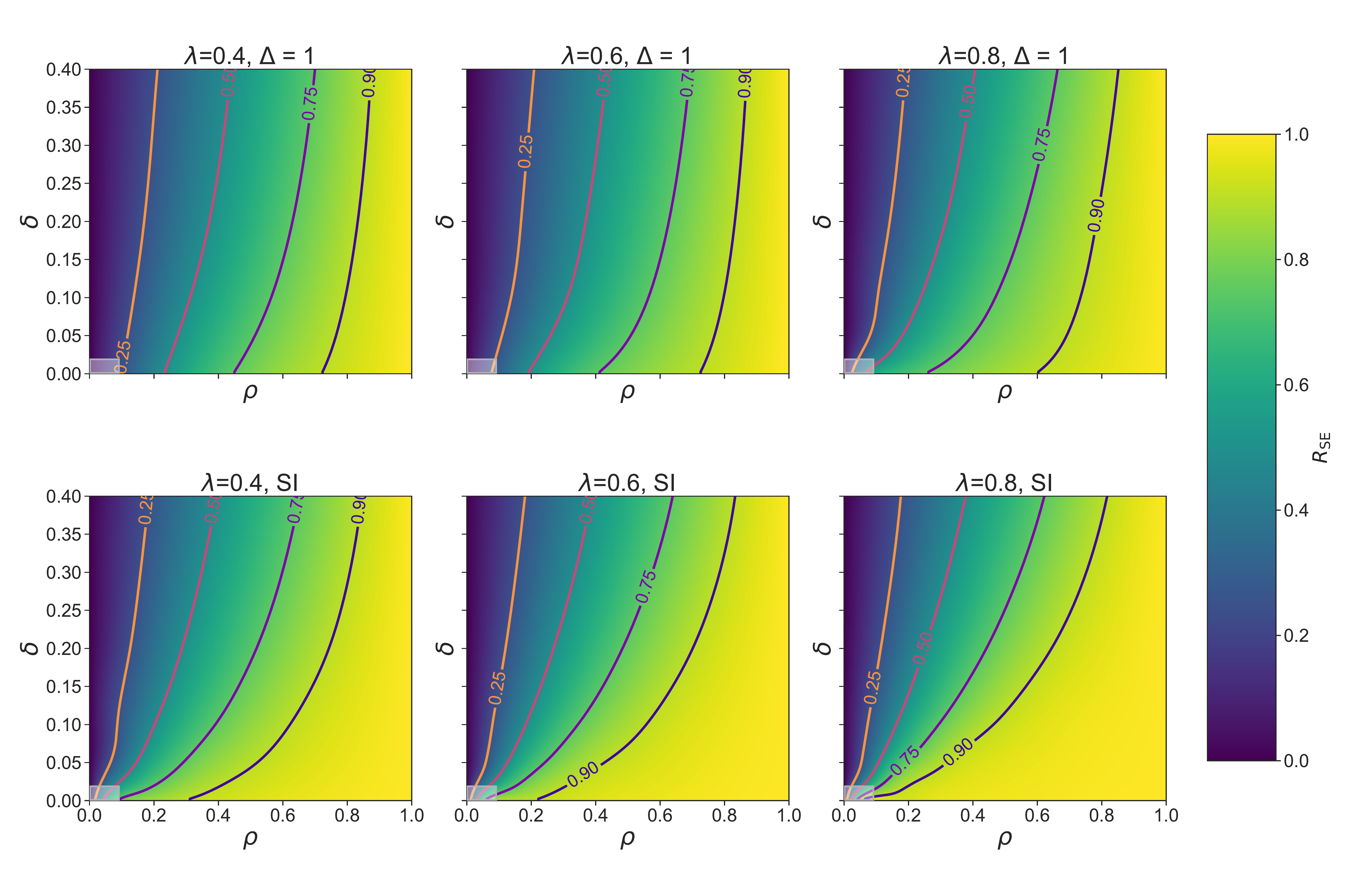}
    \caption{\label{fig:PhDiagRSE}\textbf{Phase diagrams for inferring infection times via sensors.} We show the rescaled squared error defined in Eq.~(\ref{eq:RSE}) for 3-RRGs with $N=10^4$. We juxtapose the SI model to the dSIR model with $\Delta=1$. For both, we compare three different values of $\lambda$, showing how the performance parameter varies with the fraction of sensors $\rho \in[0,1]$ and the fraction of sources $\delta\in[2.5\cdot 10^{-3},0.4]$. As described in Fig.~\ref{fig:PhDiag_Ot}, we show the results of random initialization, and the grey region of each panel corresponds to the violation of the Nishimori condition and the failure, in that region, of the BP equations to obtain Bayes Optimal results. The plots were obtained by fitting the value of the rescaled SE over a $21\times24$ data grid, averaging over 25 simulations for each point on the grid.}
\end{figure}

Another property we are interested in characterizing is the ability to infer the single times of infection for each individual in the network. Having defined the rescaled squared error in Eq.~(\ref{eq:RMSE}), in Fig.~\ref{fig:PhDiagRSE} we show its behaviour for the same set of parameters of previous results.
As we can see from the figure, the behaviour as a function of $\lambda$ is qualitatively very similar to the one shown in Fig.~\ref{fig:PhDiag_Ot} for the overlap, even if the two quantities describe two very different aspects of inference. In this case, the MSE is influenced much less than the overlap at time $0$ by the same change in the infectivity parameter $\lambda$. Furthermore, in contrast to the previous plot, now we see that the parameter $\Delta$ plays a major role in determining the value of the performance parameter: comparing the upper and lower panels of Fig.~\ref{fig:PhDiagRSE}, we see that, again especially for low values of $\delta$, is much easier (in terms of needed observations) to have a high value of the performance parameter in the case of the SI model compared to the dSIR model with $\Delta=1$. 

The great variability of the $R_{SE}$ between the case of dSIR with $\Delta=1$ and $\Delta = \infty$ (the SI model) shown in Fig.~\ref{fig:PhDiagRSE} leads us to study in detail how the performances change, varying the recovery time $\Delta$. The results are shown in Fig.~\ref{fig:OTdel} in the Appendix~\ref{app:AddImag}.

\subsection{Phase diagrams for observations via a snapshot}
In this section, we present the phase diagrams of the rescaled overlap and squared error for snapshot inference problems. In this case, the state (Susceptible or Infected) of all individuals at a given time of observation ($T_{obs}$) is probed. To link with previous literature using snapshot observations, see for instance  \cite{BIofEpi, zhu2017catch}, in the following, we consider tests that are not able to distinguish between infectious (I) and recovered individuals (R), but just to signal that the time of infection happened before the time of observation. In practice, this observation excludes all values of $t_i\geq T_{\rm obs}$. Of course, considering tests capable of distinguishing between the $I$ and the $R$ states could improve the algorithm's performance on both tasks we considered in our work.

Furthermore, a second choice we made for the analysis was to focus on the case of ``backward inference'', neglecting the study on the ability of our algorithm to infer what happens to the system after we take the snapshot. In practice, this is done simply by fixing the final time $T$ (as defined in Appendix~\ref{App:BPEPI}) to be equal to $T_{\rm obs}$.
 
 \subsubsection{Mean overlap at time zero}
 Let us start, as before, with the task of inferring the sources of epidemics, considering again for simplicity an ensemble of $3$-RRGs and comparing performances on the SI model to the ones on the dSIR model with delay $\Delta=1$. Fixing three different values of $\lambda$, we vary the fraction of sources $\delta$ and the time of observation $T_{\rm obs}$ inside an interesting interval. 
\begin{figure}
    \centering
    \includegraphics[width=0.85\textwidth]{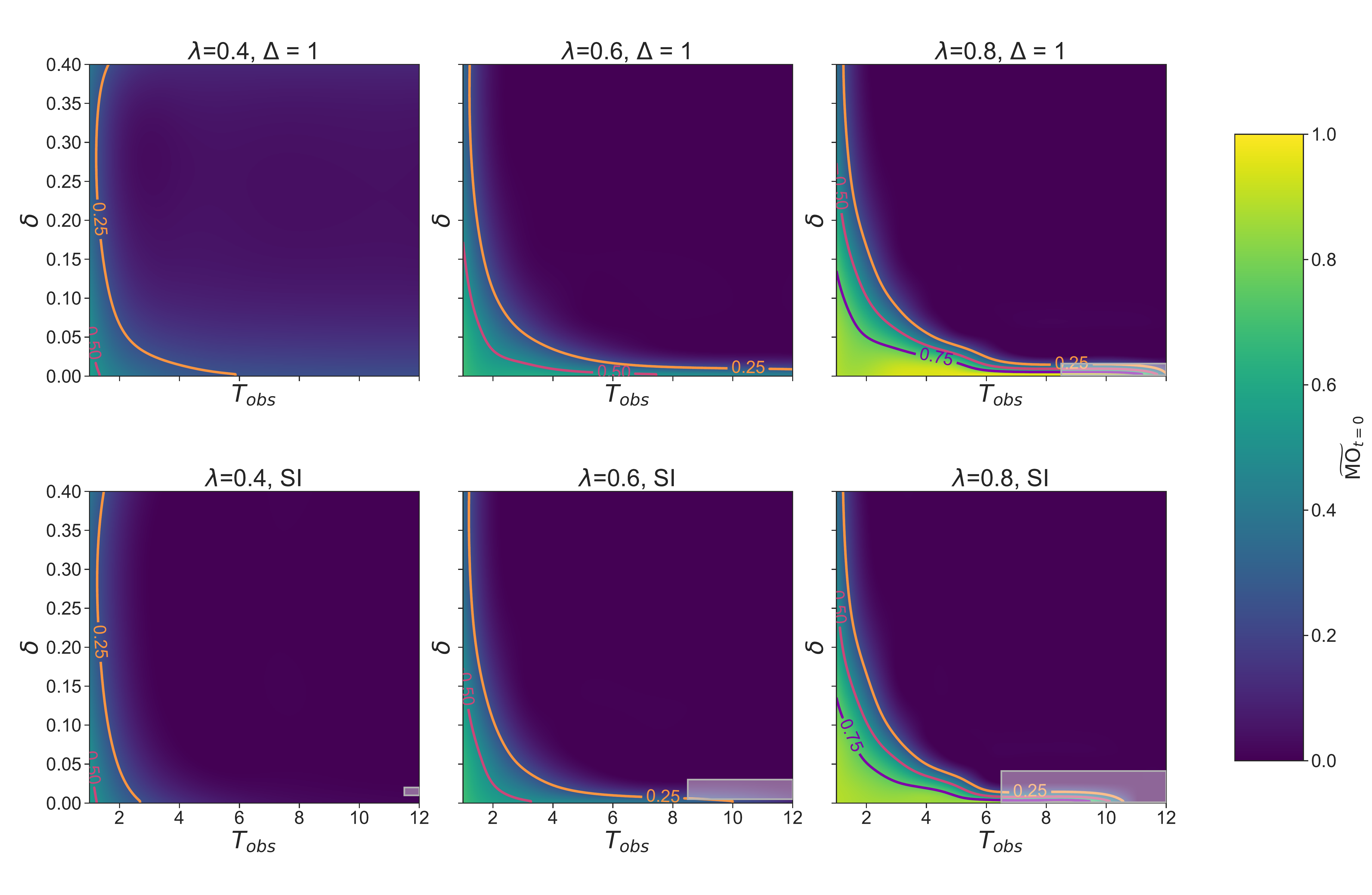}
    \caption{\label{fig:PhDiag_Ot_snap}\textbf{Phase diagrams for source-finding in the snapshot framework.} We show the rescaled mean overlap defined in Eq.~(\ref{eq:ResMOv}) for 3-RRGs with $N=10^4$, computed at time $t=0$. We juxtapose the SI model to the dSIR model with $\Delta=1$ and for both, we compare three different values of $\lambda$ and we show how the performance parameter varies with the time of observation $T_{\rm obs}\in[1,12]$ and the fraction of sources $\delta\in[2.5\cdot 10^{-3},0.4]$. As for previous phase diagrams, we show just the case of random initialization. For all the points outside the grey rectangles, the Nishimori conditions were verified, and thus also $\widetilde{\text{O}}_{t=0}$ has the same behaviour. Each contour plot was obtained by fitting the value of the rescaled mean overlap over a $12\times20$ grid of data, doing 25 simulations for each point on the grid. }
\end{figure}
The results for the rescaled mean overlap, computed using the BP algorithm and expressed by Equation (\ref{eq:ResOv}), are presented in Figure~\ref{fig:PhDiag_Ot_snap}. Upon comparison with the results for sensors shown in Figure~\ref{fig:PhDiag_Ot}, it is evident that, in general, the inference is much more challenging in the snapshot case. This is because snapshots provide less information about the dynamics than the sensor case, which reveals a finite fraction of nodes with precisely known infection times.

The results in Figure~\ref{fig:PhDiag_Ot_snap} show that increasing the observation time $T_{\rm obs}$ significantly reduces the algorithm's ability to retrieve sources. Moreover, it is possible to achieve a better-than-random performance only when the observation time is sufficiently small and $\delta$ is low. 

As for the spreading parameter $\lambda$ and the recovery delay $\Delta$, the qualitative behaviour is similar to that observed in the sensor case. Specifically, the inference of patient zero depends only slightly on $\Delta$, whereas the infectivity parameter has a more significant impact. For instance, at $\lambda=0.4$, even at small values of $T_{\rm obs}$ and $\delta$, the task is extremely challenging.

Similar to the case of observation via sensors, we identified a region in which BP has difficulty converging, and the Nishimori conditions are not met. This region is marked in a lighter colour. Once again, this occurs for small values of $\delta$ and coincides with the transition between as-random and better-than-random performance. In this case, the transition occurs when $T_{\rm obs}$ approaches the time $T$ at which the epidemic ``stops'' (See Section~\ref{sec:Nishi} for more information).
\subsubsection{Mean squared error on the times of infection\label{sec:MSEsnap}}
\begin{figure}
    \centering
    \includegraphics[width=0.85\textwidth]{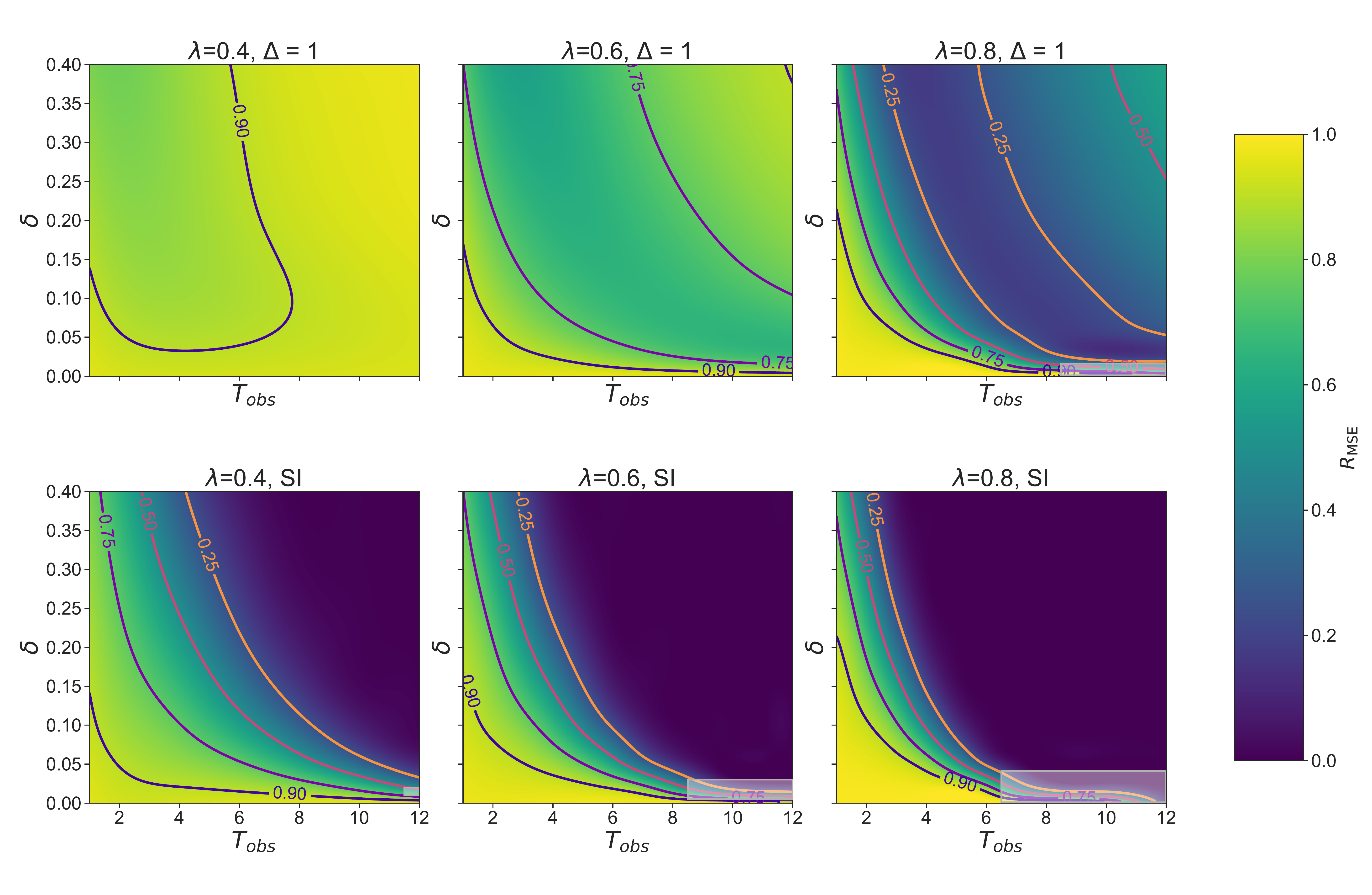}
    \caption{\label{fig:PhDiag_RSE_snap}\textbf{Phase diagrams for the inference of infection times in the snapshot framework.} We show the rescaled mean squared error defined in Eq.~(\ref{eq:RMSE}) for 3-RRGs with $N=10^4$. We juxtapose the SI model to the dSIR model with $\Delta=1$ and for both, we compare three different values of $\lambda$ and we show how the performance parameter varies with the time of observation $T_{\rm obs} \in[1,12]$ and the fraction of sources $\delta\in[2.5\cdot 10^{-3},0.4]$. As for previous phase diagrams, we show just the case of random initialization. For all the points outside the grey rectangles, the Nishimori conditions were verified, and thus also $R_{SE}$ has the same behaviour. Each contour plot was obtained by fitting the value of the rescaled MSE over a $12\times20$ grid of data, doing 25 simulations for each point on the grid.}
\end{figure}

In the following, we analyse the performances to infer the entire time trajectory of the individuals, measured in terms of the rescaled mean squared error $R_{\rm MSE}$, Eq.~(\ref{eq:RSE}). The results are presented in Fig.~\ref{fig:PhDiag_RSE_snap}, where we consider both the SI model and the dSIR model with $\Delta=1$.

Interestingly, we achieve much better performance in inferring the time trajectory of the nodes than those obtained for the source-finding task, with the same values of epidemic parameters. Specifically, we can always find a region in which $R_{\rm SE}>0.9$ for $\lambda\geq0.4$.

As one would expect, for the SI model, we see that, for any arbitrary $\delta$, increasing $T_{\rm obs}$ implies a decrease in performance. This is explained by the fact that a snapshot done later in time brings very little information on the epidemic compared to snapshots performed earlier in time. The limit case is doing the snapshot after the epidemic has stopped when all nodes are infected, and we cannot do better than the random estimator, resulting in $R_{\rm MSE}=0$.

The dSIR model displays a different behaviour, as is evident by looking at the upper panels of Figure~\ref{fig:PhDiag_RSE_snap}. At fixed  arbitrary $\delta$ and increasing $T_{\rm obs}$, we observe, in the beginning, a decrease in performance (as in the SI model). But after reaching a minimum, the $R_{\rm MSE}$ increases. This is because for the dSIR model, at the end of the epidemic, a fraction of the nodes are recovered (R), and the rest are still susceptible (S). This implies that by doing a snapshot after the epidemic is finished, we still retain the information on who is still susceptible. This explains why the value of $R_{\rm MSE}$ does not go to zero as $T_{\rm obs}$ increases. To explain the non-monotonicity, we must look at the definition of the $R_{\rm MSE}$ in Eq.~(\ref{eq:RMSE}), where the $\text{\rm MSE}$ is rescaled with the $\text{\rm MSE}(\hat{\mathbf{t}}^{\text{RND}})$, the random estimator. This last observable gets worse as we increase $T_{\rm obs}$ because the possible times of infections increase as well. The result is the observed non-monotonicity of the $R_{\rm MSE}$ as a function of the time of observations $T_{\rm obs}$. As an illustration of this phenomenon, in Fig.~\ref{fig:SEvsRND} in the Appendix \ref{app:AddImag} we fix $\lambda=0.6$, and some values of $\delta$ for which this behaviour is clearly visible by looking at SE and $\text{SE}(\hat{\mathbf{t}}^{\text{RND}})$ for the case $\Delta=1$ and $\Delta=\infty$.

Similar arguments allow us to explain the impact of the infectivity parameter $\lambda$ on the performance of BP. We can notice that decreasing $\lambda$ increases the number of susceptible individuals at the end of the simulation. In turn, their times of infection will be inferred perfectly thanks to the information obtained from the snapshot.
\section{\label{sec:Nishi}Regions with failure of Bayes-optimality}

The previous section presented the performance of the Belief Propagation (BP) algorithm in two different settings of inference. We showed large regions of parameters where the BP equations converge and satisfy the Nishimori conditions. We thus conjecture that in these regions, the solutions found by the BP equations are very close to the Bayes-optimal performance.

\begin{figure}
    \centering
    \includegraphics[width=\textwidth]{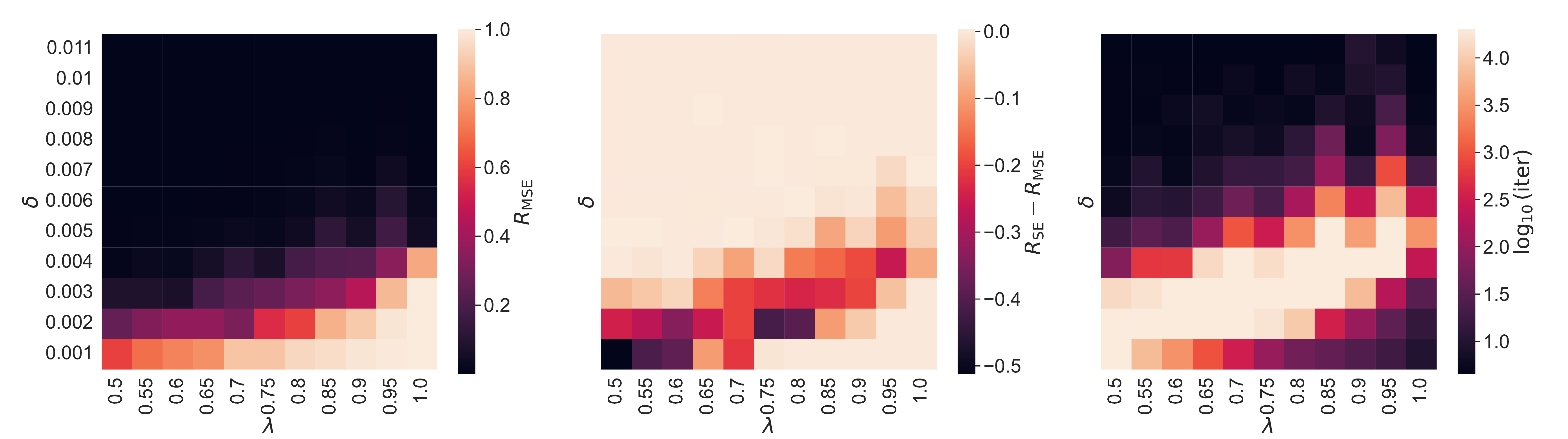}
    \caption{\textbf{Failure of the Nishimori conditions.} Example of inference for snapshot observations in a region of parameters where Bayes-optimality fails. The plots were generated using an ensemble of 3-regular random graphs (3-RRGs) with $N=10^4$ and varying values of $\lambda$ and $\delta$. We consider an SI model where the snapshot time is fixed at $T_{\rm obs}=9/\lambda$. In the left panel, we show the performance of BP through the rescaled MSE defined in equation~(\ref{eq:RMSE}). The central panel probes the Nishimori conditions by studying how $\delta R_\text{SE} \equiv R_\text{SE} - R_\text{\rm MSE}$ behaves in the same range of parameters. The right panel shows the number of iterations for BP to converge. We set the maximum number of iterations to $2\cdot 10^4$, so that the bright points indicate instances where BP did not converge. Each point was averaged over ten instances.}
    \label{fig:NishiSnapPD}
\end{figure}

It should be noted that observing such behaviour at relatively small sizes is rather surprising. The mathematical arguments about the exactness of BP rely on the fact that loops in the graph are longer than the correlation length. But since the length of the loops only grows logarithmically with size, this condition seems unreasonable to graphs of size as we treat in our experiments. Yet, we observe that BP converges and the Nishimori conditions are satisfied. Similar surprisingly small finite-size effects have been observed in many previous works using BP. Sometimes already at the size of several thousand, we see behaviour that is very close to the predicted thermodynamic limit. It is not clear why this is so. 

As pointed out in Sec.~\ref{sec:BOresults}, we observed narrow regions of parameters (marked in clear colour in the phase diagrams) where the BP equations do not converge, and the Nishimori conditions are no longer satisfied. 
For the two observation scenarios considered, the sensor and snapshot ones, we observe that the regions where BP fails are always at small values of $\delta$ (the fractions of sources) and during the transition from regions where inference is impossible, i.e., the optimal solutions found by BP equals those of the random estimator, and regions where a close to perfect recovery of the missing information is achieved. 
As an example, we will focus on the case of snapshot observations, even though a similar behaviour can be observed for sensors.

We investigate these regions closely and conclude that what is observed is not a sign of a critical region (where one would get a diverging length scale) and, thus, must be a finite-size effect. However, for the sizes we can simulate, this trouble does not go away. The existence of this region should perhaps not be surprising; what should be surprising is the region presented previously, where even for very moderate sizes, the thermodynamic limit is already effectively reached. We did not identify anything crisp, but we do hope that pointing to the existence of a region where BP should be asymptotically optimal but for the considered rather large sizes it is not will help to shed some light on the fact that BP is often able to perform well and converge even when the loops are still very short. 

In Fig.~\ref{fig:NishiSnapPD}, we investigate the inference from a snapshot at a certain time $T_{\rm \rm \rm obs}$, and then we infer the infection times before $T_{\rm \rm obs}$, disregarding the subsequent epidemic evolution. We consider different values of $\lambda\in[0.5,1]$ and we rescale the snapshot time as 
\begin{equation}
    T_{\rm \rm obs}(\lambda) = \frac{T_{\rm \rm obs}(1)}{\lambda}\,,
\end{equation}
 to ensure that we observe (on average) the same ``epidemic time'', i.e., keeping roughly constant the fraction of susceptible nodes at the time of observation. The left panel of Fig.~\ref{fig:NishiSnapPD} shows the values of the $R_{\rm MSE}$ varying $\lambda$ and $\delta$. The central and right panels, where we investigate the Nishimori conditions and the number of iterations needed for BP to converge, show that during the transition regions of the $R_{\rm MSE}$, the algorithm struggles to converge, and the Nishimori conditions are no longer satisfied. 

\begin{figure}
    \centering
    \includegraphics[width=\textwidth]{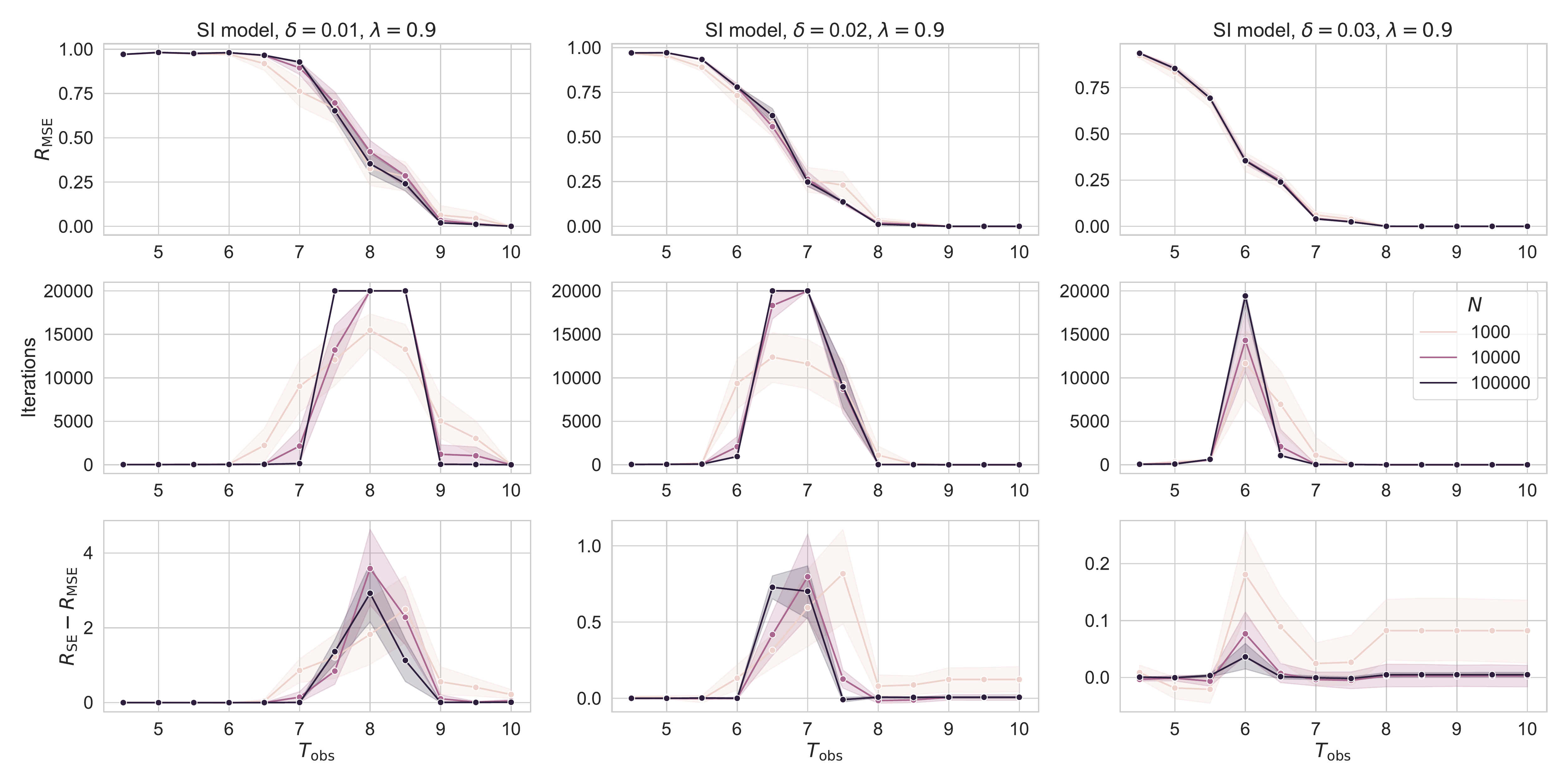}
    \caption{\textbf{Finite size scaling.} The plots were generated simulating an SI model on 3-regular random graphs (3-RRGs) of varying size, using snapshots at times $T_{\rm \rm obs}$ as observations and starting BP with uninformed messages. The fraction of source nodes varies $\delta=0.01, 0.02, 0.03$. The first row shows the values of $R_{\rm MSE}$ computed with the BP equations and averaged over 20 instances. The plots in the second row display the number of iterations to reach convergence, with a maximum of allowed iterations of $2\cdot 10^4$. The plots in the third row show the difference $R_{SE} - R_{\rm MSE}$, which certifies the violation of the Nishimori conditions when different from zero. We observe that the region of violation of the Nishimori condition may shrink very slowly with the system size. }
    \label{fig:Scaling_nishi}
\end{figure}

Fig.~\ref{fig:Scaling_nishi} shows a finite-size effect study of the non-convergence region, again in the snapshot case. The values of the observation time $T_{\rm \rm obs}$, the infectivity parameter $\lambda$, and the fraction of sources $\delta$ is chosen to observe the transition where close to perfect inference is possible to regions where we obtain the same results of a random estimator.
First, we observe that the values of the observable $R_{\rm MSE}$ (first-row of Fig.~\ref{fig:Scaling_nishi}) basically do not change with $N$, the number of nodes considered. The curve at $N=10^4$ and $N=10^5$ are almost indistinguishable, comforting the choice of $N=10^4$ as the primary size used to characterize the phase space of the inference problems analysed. The small values of $\delta$ chosen allow us to observe, during the transition on the $R_{\rm MSE}$ values, that the BP equations stop to converge, the second row of Fig.~\ref{fig:Scaling_nishi}. Observing the violation of the Nishimori conditions, the third row of Fig.~\ref{fig:Scaling_nishi}, we can note that the regions where they are violated shrink very mildly. This fact confirms our hypothesis that the non-converge of the BP equations and the violation of the Nishimori conditions may be due to finite-size effects problems; nevertheless, the improvement of the Nishimori condition is very slow with $N$, making the BP equation not Bayes-optimal in practical ranges of system sizes. For instance, this is the case of the very-well known patient-zero problem, where only one source infection is supposed to start the diffusion process. 

\begin{figure}[t!]
    \centering
    \includegraphics[width=0.9\textwidth]{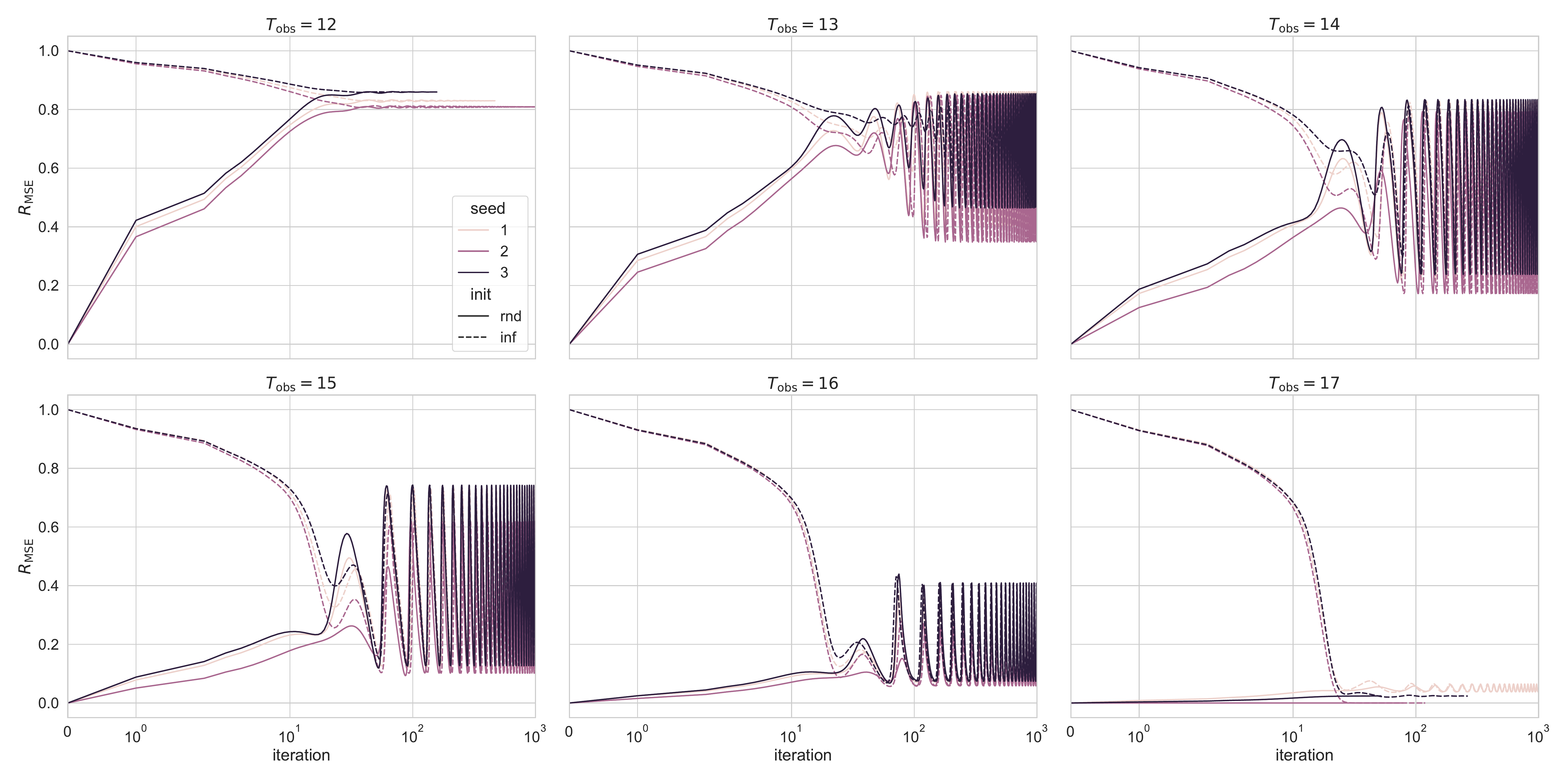}
    \caption{\textbf{Absence of critical slowing down.} The plots were generated using an ensemble of 3-regular random graphs (3-RRGs) with $N=10^4$, $\lambda = 0.5$ and $\delta=0.005$. We consider an SI model where the snapshot time is changed in each panel between $T_{\rm obs}=12$ (top left), and $T_{\rm obs}=17$ (bottom right). For every panel, we examine three instances of the problem generated randomly, and we monitor the rescaled mean squared error (\ref{eq:RMSE}) during the BP iterations. We compare the results of the same realizations when we initialize the messages, whether informed or not. The damping parameter in the BP algorithm was fixed to $\eta = 0.4$.}
    \label{fig:Noslow}
\end{figure}

We also investigate whether the observed issue cannot be a sign of proximity to a critical point. 
In Figure~\ref{fig:Noslow}, we consider again the SI model with snapshot observations, fixing $\lambda=0.5$ and $\delta=0.005$ and considering values of $T_{\rm obs}$ around the problematic region, which in this case is observed to be around $T_{\rm obs}=14$. In the presence of a critical point, one should observe what in statistical physics is called ``critical slowing down'', i.e., a power law increase in convergence time when approaching the critical point. Since the algorithm does not converge in our case, we track the two initializations separately. We study the iteration time for both initializations to reach approximately the same value. Fig.~\ref{fig:Noslow} shows the $R_{\rm MSE}$ values in the intervals from $T_{\rm obs}=12$, where we take the snapshot early enough to make a good inference, and $T_{\rm obs}=17$, where conversely we do it late enough to have no added information with respect to a random estimator. The time needed for the two initialization to have similar values does not present a significant change. The same behaviour can be observed by looking directly at the difference between BP messages, as shown in Fig.~\ref{fig:SEmess} in App.~\ref{app:AddImag}.

Finally, after reaching the point where the two initializations predict almost the same observable value, both  oscillate. The effect is present even with random updates of the BP equations and several different values of the damping parameter; see Appendix \ref{App:BPEPI} for details about the damping parameter definition.

\section{Conclusions}
In this paper, we present a detailed analysis of the performance of the Belief Propagation algorithm for spreading models defined on random networks. Compared to previous work on this topic, we have focused here on investigating whether such an algorithm is Bayes-optimal, and on the possible presence of phase transitions in the relative phase diagrams. This is done by analysing the convergence of BP from both the random and informed initial conditions, looking for first-order phase transitions. We find no such phase transitions. On the other hand, we analyse consistency conditions known as Nishimori conditions in the domain of disordered systems, which allowed us to heuristically investigate Bayes optimality in our simulations. When these conditions are satisfied, we conjecture that Belief Propagation asymptotically achieves the Bayes-optimal performance. We have provided phase diagrams both for the classical task of finding the sources of an epidemic (given different types of information) and for characterizing the entire trajectory of the nodes' state.

Furthermore, we dedicate the last section to describe our efforts in understanding the algorithm's behavior in the particular regime when the failure of the Nishimori conditions and the lack of convergence tells us we are not Bayes optimal. Although not reaching a firm conclusion, we feel confident in ruling out the hypothesis of the presence of some kind of phase transition, and we attribute this kind of behaviour to finite-size effects leading to spurious correlations in the graphs that would disappear very slowly (e.g. logarithmically) with the system size. 

\section*{Acknowledgments}
We thank Patrick Thiran, Florent Krzakala, Alejandro Lage Castellanos, and Laurent Massoulié for useful  discussions.

\newpage
\appendix
\section*{Appendix}
\section{\label{App:BPEPI} BPEpI}

\subsection{BP implementation for the dSIR model}
The dynamical process of a generic dSIR model can be described using a single time of infection $\{t_i\}_{i=1}^{N}$ for each variable, where $t_i \in \{-1,0,1,\dots,T-1,T\}$ such that:
\begin{itemize}
    \item $t=-1$ is the (fictitious) time in which the sources get infected,
    \item $t=0$ is the first ``simulation time'', i.e. the first time in which the rest of the individuals can get the infection from the sources,
    \item $t=T-1$ is the last simulation time, i.e. the last time an individual can get the infection from a neighbour,
    \item $t=T$ is the (fictitious) time in which, conventionally, all the individuals that have not been infected until (and at) time $t=T-1$, get infected\footnote{Notice that these nodes are still susceptible at time $T$, thus this does not affect observables computed at $t=T$.}.
\end{itemize}

It is not hard to see that the model just defined can be described with the formalism introduced in Section~\ref{sec:BaySet}. We define the probability of the susceptible node $i$ not to be infected at time $s$ as $p_{SS}(s,  \{t_k\}_{k\in\partial_i}) = \prod_{k\in \partial_i} (1-\lambda_{ki}(s-t_k)\Ids[s>t_k])$, and in the following, we will focus on the case in which the probability to be a source of the epidemic is uniform among the individuals, such that $P(t_i = -1) = \delta\;\forall\,i$. The factor of the posterior probability entering in  the BP equations~(\ref{eq:BP_gen}) can be written as:
\begin{equation}\label{eq:Psi_dSIR}
\begin{split}
     \widetilde{\Psi}_i\left(t_i, \{t_k\}_{k\in\partial_i}| \bm{\lambda}\right) = \Bigg[ &(1-\delta)\left( \prod_{s=0}^{t_i-1}  p_{SS}(s,  \{t_k\}_{k\in\partial_i})\right)\left( 1 - p_{SS}(t_i,  \{t_k\}_{k\in\partial_i}) \right)\Ids[-1<t_i<T] + \\ &+ (1-\delta)\left( \prod_{s=0}^{T-1}  p_{SS}(s,  \{t_k\}_{k\in\partial_i})\right)\Ids[t_i = T] + \delta\Ids[t_i=-1]\Bigg]P(\mathcal{O}_i|t_{i}).
\end{split}
\end{equation}
The first term in the square brackets represents the probability of the node $i$ to be infected at time $t_i$ by one or more of its neighbours, while the second term is the probability to remain susceptible during the whole epidemic process, and the third one is the probability to be a source of the epidemic. Inserting the above factor in Eq.~(\ref{eq:BP_gen}) we observe easily that the sum over the times of infections of neighbours can be factorized, and we now show that the time complexity of the algorithm is reduced to $\mathcal{O}(E*d_{max}T^2)$, thus no more exponential in the maximum degree of the interaction network and only quadratic in the maximum time $T$. 

In this appendix, we go into the details of the belief propagation equations, explaining how the code~\cite{bpepi_github} is implemented for the deterministic-SIR model.

Let's start by considering again the factor contribution~(\ref{eq:Psi_dSIR}) and noticing that we can rewrite the product
$$
\left( \prod_{s=0}^{t_i-1}  p_{SS}(s,  \{t_k\}_{k\in\partial_i})\right)\left( 1 - p_{SS}(t_i,  \{t_k\}_{k\in\partial_i}) \right)
$$
as
\begin{equation}
    \prod_{k\in\partial_i}  \prod_{s=0}^{t_i-1} (1-\lambda_{ki}(s-t_k))^{\theta(s-t_k)} - \prod_{k\in\partial_i} \prod_{s=0}^{t_i}(1-\lambda_{ki}(s-t_k))^{\theta(s-t_k)}\,.
\end{equation}
Then, putting it back into Eq.~(\ref{eq:BP_gen}), we get, apart from normalization
\begin{equation}\label{eq:BPepi_mess}
\begin{split}
    m_{i\rightarrow j}(t_i,t_j) \propto &(1-\delta)P(\Ocl_i|t_i)\Ids[-1<t_i<T]\times \\ &\times\Bigg[ \Big( \prod_{s=0}^{t_i-1} (1-\lambda_{ji}(s-t_j))^{\theta(s-t_j)} \Big)  \prod_{k\in\partial_i\backslash j} \sum_{ t_k = -1}^T \Big[\Big( \prod_{s=0}^{t_i-1} (1-\lambda_{ki}(s-t_k))^{\theta(s-t_k)} \Big)  m_{k\rightarrow i} (t_k,t_i)\Big] - \\ &- \Big( \prod_{s=0}^{t_i}(1-\lambda_{ji}(s-t_j))^{\theta(s-t_j)} \Big)  \prod_{k\in\partial_i\backslash j} \sum_{ t_k = -1}^T \Big[\Big( \prod_{s=0}^{t_i}(1-\lambda_{ki}(s-t_k))^{\theta(s-t_k)} \Big)  m_{k\rightarrow i} (t_k,t_i)\Big]\Bigg]+ \\ &+ (1-\delta)\Ids[t_i=T]P(\Ocl_i|t_i=T)\Big( \prod_{s=0}^{T-1}(1-\lambda_{ji}(s-t_j))^{\theta(s-t_j)} \Big) \times \\ &\times  \prod_{k\in\partial_i\backslash j}\sum_{t_k=-1}^T\left( \prod_{s=0}^{T-1} (1-\lambda_{ki}(s-t_k))^{\theta(s-t_k)}\right)   m_{k\rightarrow i} (t_k,T) + \\ &+ \delta\Ids[t_i=-1] P(\Ocl_i|t_i=-1) \prod_{k\in\partial_i\backslash j}\sum_{ t_k = -1}^T m_{k\rightarrow i} (t_k, t_i=-1)\,.
\end{split}
\end{equation}
\subsection{Practical implementation}
We define the following matrices for each directed edge $i,j$:
\begin{align*}
    \Lambda^1_{ji}\left(t_j,t_i\right) &\equiv \prod_{s=0}^{t_i-1} (1-\lambda_{ji}(s-t_j))^{\theta(s-t_j)}\\
    \Lambda^0_{ji}\left(t_j,t_i\right) &\equiv \prod_{s=0}^{t_i} (1-\lambda_{ji}(s-t_j))^{\theta(s-t_j)}
\end{align*}
where one can easily see that $\Lambda^0_{ij}\left(t_i,t_j\right) =  \Lambda^1_{ij}\left(t_i,t_j\right) * (1-\lambda_{ji}(t_i-t_j)))^{\theta(t_i-t_j)}$. The messages read:
\begin{equation}
\label{eq:BPmess}
\begin{split}
    m_{i\rightarrow j}(t_i,t_j) \propto &(1-\delta)P(\Ocl_i|t_i)\Ids[-1<t_i<T] 
    \times  \Bigg[ \Lambda^1_{ji}\left(t_j,t_i\right)
    \prod_{k\in\partial_i\backslash j} \sum_{ t_k = -1}^T \Big[\Lambda^1_{ki}\left(t_k,t_i\right) m_{k\rightarrow i} (t_k,t_i)\Big] + \\ &- \Lambda^0_{ji}\left(t_j,t_i\right)
    \prod_{k\in\partial_i\backslash j} \sum_{ t_k = -1}^T \Big[\Lambda^0_{ki}\left(t_k,t_i\right)  m_{k\rightarrow i} (t_k,t_i)\Big]\Bigg]+ \\ &+ (1-\delta)\Ids[t_i=T]P(\Ocl_i|t_i=T)\Lambda_{ji}^1(t_j,T) \prod_{k\in\partial_i\backslash j} \sum_{ t_k = -1}^T \Big[\Lambda^1_{ki}\left(t_k,T\right) m_{k\rightarrow i} (t_k,T)\Big] + \\ &+ \delta \Ids[t_i=-1]P(\Ocl_i|t_i=-1) \prod_{k\in\partial_i\backslash j}\sum_{ t_k = -1}^T m_{k\rightarrow i} (t_k, t_i=-1)\,,
\end{split}
\end{equation}
that can also be rewritten as:
\begin{equation}
\label{eq:msgs_imp}
\begin{split}
    m_{i\rightarrow j}(t_i,t_j) \propto &(1-\delta)\Ids[-1<t_i<T]P(\Ocl_i|t_i) 
    \times  \Bigg[ \Lambda^1_{ji}\left(t_j,t_i\right)
    \prod_{k\in\partial_i\backslash j} \gamma^1_{ki}(t_i) - \Lambda^0_{ji}\left(t_j,t_i\right)
    \prod_{k\in\partial_i\backslash j} \gamma^0_{ki}(t_i) \Bigg]+ \\ &+ (1-\delta)\Ids[t_i=T]P(\Ocl_i|t_i=T)\Lambda_{ji}^1 (t_j,T) \prod_{k\in\partial_i\backslash j} \gamma^1_{ki}(T) + \\ &+ \delta\Ids[t_i=-1] P(\Ocl_i|t_i=-1) \prod_{k\in\partial_i\backslash j}\sum_{ t_k = -1}^T m_{k\rightarrow i} (t_k, t_i=-1)\,,
\end{split}
\end{equation}
where we have defined the quantities:
\begin{align*}
\gamma^1_{ki}(t_i) &\equiv \sum_{ t_k = -1}^T \Big[\Lambda^1_{k,i}\left(t_k,t_i\right)  m_{k\rightarrow i} (t_k,t_i)\Big] \\
\gamma^0_{ki}(t_i) &\equiv \sum_{ t_k = -1}^T \Big[\Lambda^0_{k,i}\left(t_k,t_i\right)  m_{k\rightarrow i} (t_k,t_i)\Big] 
\end{align*}
The marginals for the infection times $t_i$ can be obtained, after fixing as $j$ one neighbour of $i$, through the following equation:
\begin{equation}
    b_i(t_i) \propto \sum_{t_j=-1}^{T} m_{i\rightarrow j}(t_i, t_j)m_{j\rightarrow i}(t_j, t_i)
\end{equation}
The belief propagation algorithm is implemented in Python and can be found in~\cite{bpepi_github}. Additionally, in \cite{figures_github}, there is the code to reproduce every figure in the paper.
\subsubsection{\label{sec:damp}Damping}
To improve the convergence of the Belief Propagation algorithm, as the one following from Eq.~(\ref{eq:BPmess}), a damping updating scheme is typically used. In practical terms, the new messages are a linear combination of the old and new ones. In mathematical terms, writing as $m^n$ the message at iteration $n$, we use
\begin{equation}
    m^{n+1} = \eta m^n + (1-\eta)f_{\rm BP}(\bm{m^n})\,,
\end{equation}
where we write as $f_{\rm BP}(\bm{m^n})$ the r.h.s. of Eq.~(\ref{eq:BPmess}), and the parameter $\eta$ is introduced to control the intensity of the damping. 

If not specified otherwise, in the simulations in the main text, the damping was set to $0$ for the first 200 iterations, to $0.2$ for the next 200, and then $0.4$ until the end.
\subsubsection{Notes about implementation}
The messages can be stored in a NumPy tensor of shape $2|E|\times (T+2)\times (T+2)$, as well as $\Lambda^0$ and $\Lambda^1$ that can be pre-computed from the beginning. These data are stored in a Python class called \textit{FactorGraph}. We implement a class called \textit{SparseTensor} that represents the tensor of shape $2|E|\times T\times T$, and includes a transparent mapping between the pair of nodes $(i,j)$ and the corresponding labelled edge.
\paragraph{Efficient implementation of Eq.~(\ref{eq:msgs_imp})} For each node $i$ we update, at the same time, all messages that come out. We start by computing the following quantities:
\begin{itemize}
    \item $\gamma^1_{ki}(t_i)$ and $\gamma^0_{ki}(t_i)$ for all $k\in\partial i$\,,
    \item $\gamma^1_i(t_i) = \prod_{k\in \partial i}\gamma^1_{ki}(t_i)$ and $\gamma^0_i(t_i) = \prod_{k\in \partial i}\gamma^0_{ki}(t_i)$\,.
\end{itemize}
Calling $\widehat{M_{ij}}$ the matrix containing the message $m_{i\rightarrow j}(t_i,t_j)$ for each $t_i$ and $t_j$, the update is described by the following matrix equation:
\begin{align*}
    \widehat{M_{ij}}[1:T+1,:] &= (1-\delta) \underline{O_i} \cdot \left[ \widehat{\Lambda^1_{ji}}[:,1:T+1] \cdot  \underline{\gamma_i^1} \cdot \left(\underline{\gamma^1_{ji}} \right)^{-1}
    - \widehat{\Lambda^0_{ji}}[:,1:T+1] \cdot  \underline{\gamma_i^0} \cdot \left(\underline{\gamma^0_{ji}}\right)^{-1}   \right]\\
    \widehat{M_{ij}}[T+1,:] &= (1-\delta) O_i(T) \widehat{\Lambda^1_{ji}}[:,T+1] \cdot  \frac{\gamma_i^1(T)}{\gamma^1_{ji}(T)}     \\
    \widehat{M_{ij}}[0,:] &= \delta\, O_i(-1) \prod_{k\in\partial_i\backslash j} \sum_{t_k=-1}^T\widehat{M_{ki}}[t_k,0]
\end{align*}
The $\cdot$ are element-wise products, which are handled using the indexing rule of NumPy \cite{harris2020array}. $\underline{O_i}$, $\underline{\gamma_i^{1/0}}$ and $\underline{\gamma^{1/0}_{ij}}$ are $(T+2)$-dimensional vectors. $\widehat{M_{ij}}$, $\widehat{\Lambda^{1}_{ij}}$ and $\widehat{\Lambda^{0}_{ij}}$ are $(T+2)\times (T+2)$ matrices, where the first entry is $t_i$ and the second one is $t_j$.
\section{\label{app:AddImag}Additional Images}
\subsection{Inference of the final time}
\label{sec:MOTdel}
We can consider the case of what happens when we try to infer the state of the system at final time~$T$, which is trivial for the SI model since all nodes are infected, but become more interesting for the dSIR model, where a fraction of nodes remains susceptible at the end of the epidemic. We consider the dSIR model with delay parameter $\Delta$. 
\begin{figure}
    \centering
    \includegraphics[width=0.7\textwidth]{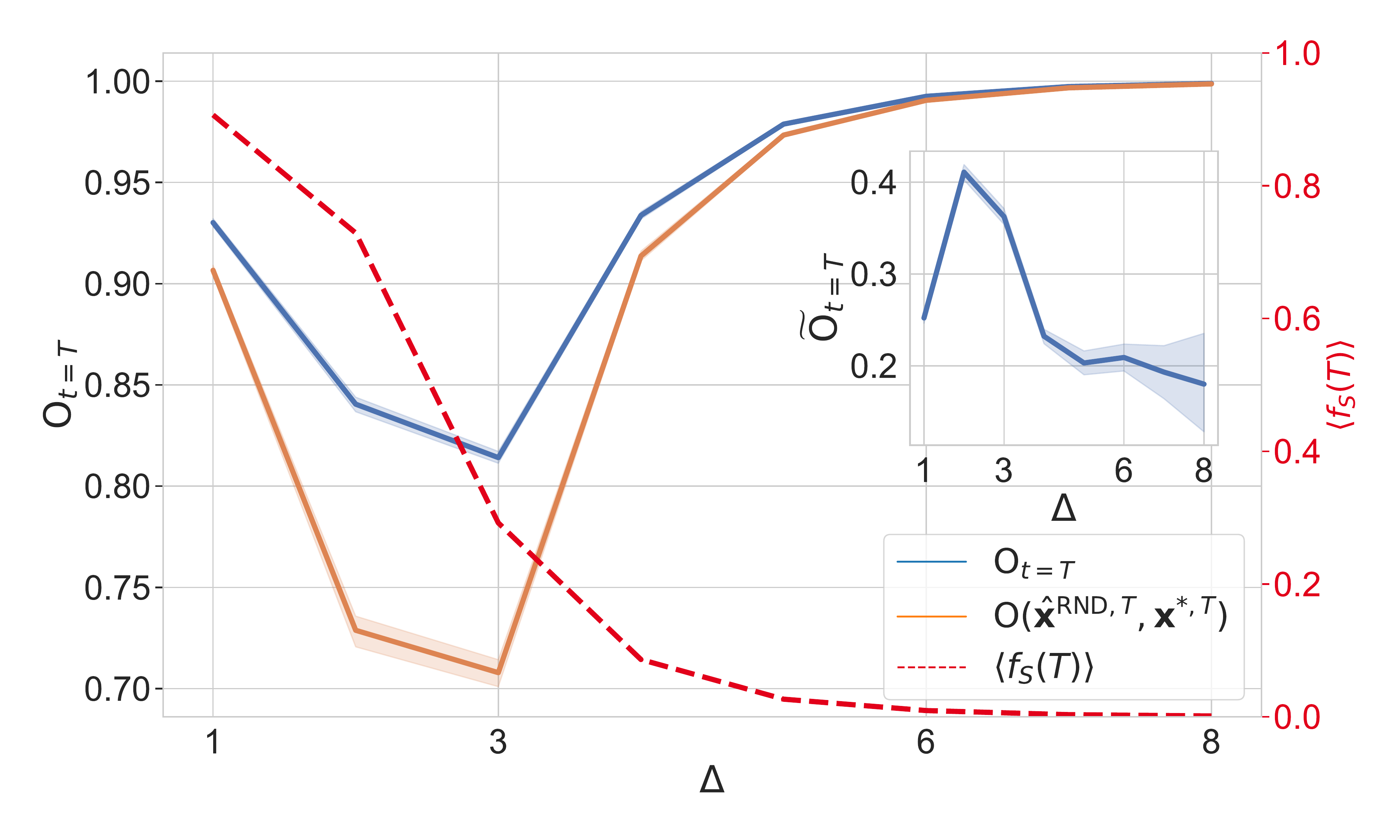}
    \caption{\label{fig:OTdel}\textbf{Inference of the state at the final time as a function of $\Delta$.} The plots were generated using an ensemble of 3-regular random graphs (3-RRGs), fixing $\lambda=0.25$ and $\delta=0.04$ for the spreading. Furthermore, we consider sensor observations, with a fraction of observed nodes $\rho=0.2$. We compare the overlap (\ref{eq:ov}) at $t=T$ computed with $\widehat{\boldsymbol{x}}^{\text{MMO}, T}$ and with $\widehat{\boldsymbol{x}}^{\text{RND}, T}$ varying $\Delta$ in a dSIR model. The dashed red line represents the average fraction of susceptible nodes at $t=T$. The inset plot represents the behavior of the rescaled overlap defined in (\ref{eq:ResOv}) for the same range of parameters. We verified that the Nishimori conditions were satisfied for all the points displayed in the plots and that the informed initialization led to the same result as the uninformed one.}
\end{figure}

In Fig.~\ref{fig:OTdel} the ability of the algorithm to infer the state of the system at the final time $T$ is shown.
The average value of the fraction of susceptible nodes at the end of the epidemic is represented as the dashed red line in the figure, and we see that, as expected, as we increase $\Delta$ we go back to the SI model, where we perform similar to the random estimator. If, instead, we take a lower value of $\Delta$, we can have less trivial situations, with the hardest one being the case in which the fraction of susceptible nodes $f_S(T)\approx 1/2$, in which the random estimator performs worse. Nevertheless, it is interesting to notice that this is also the regime where the algorithm performs best, as can be seen from the inset plot in 
Figure \ref{fig:OTdel}, in which the behavior of the rescaled overlap with $\Delta$ is displayed.

\subsection{Rescaled squared error for snapshot observations in the dSIR model}
In this section, we present an example to explain the behaviour described in Section~\ref{sec:MSEsnap} where the results of the rescaled squared error for snapshot observations were presented. In particular, concerning Figure~\ref{fig:PhDiag_RSE_snap}, we additionally fix $\lambda=0.6$, but the same behavior is visible for every value of the infectivity parameter. We then compare what happens for the SI model ($\Delta=\infty$) and for the dSIR model with $\Delta=1$, looking separately at the SE computed with the optimal estimator and with the random estimator.
\begin{figure}
    \centering
    \includegraphics[width=\textwidth]{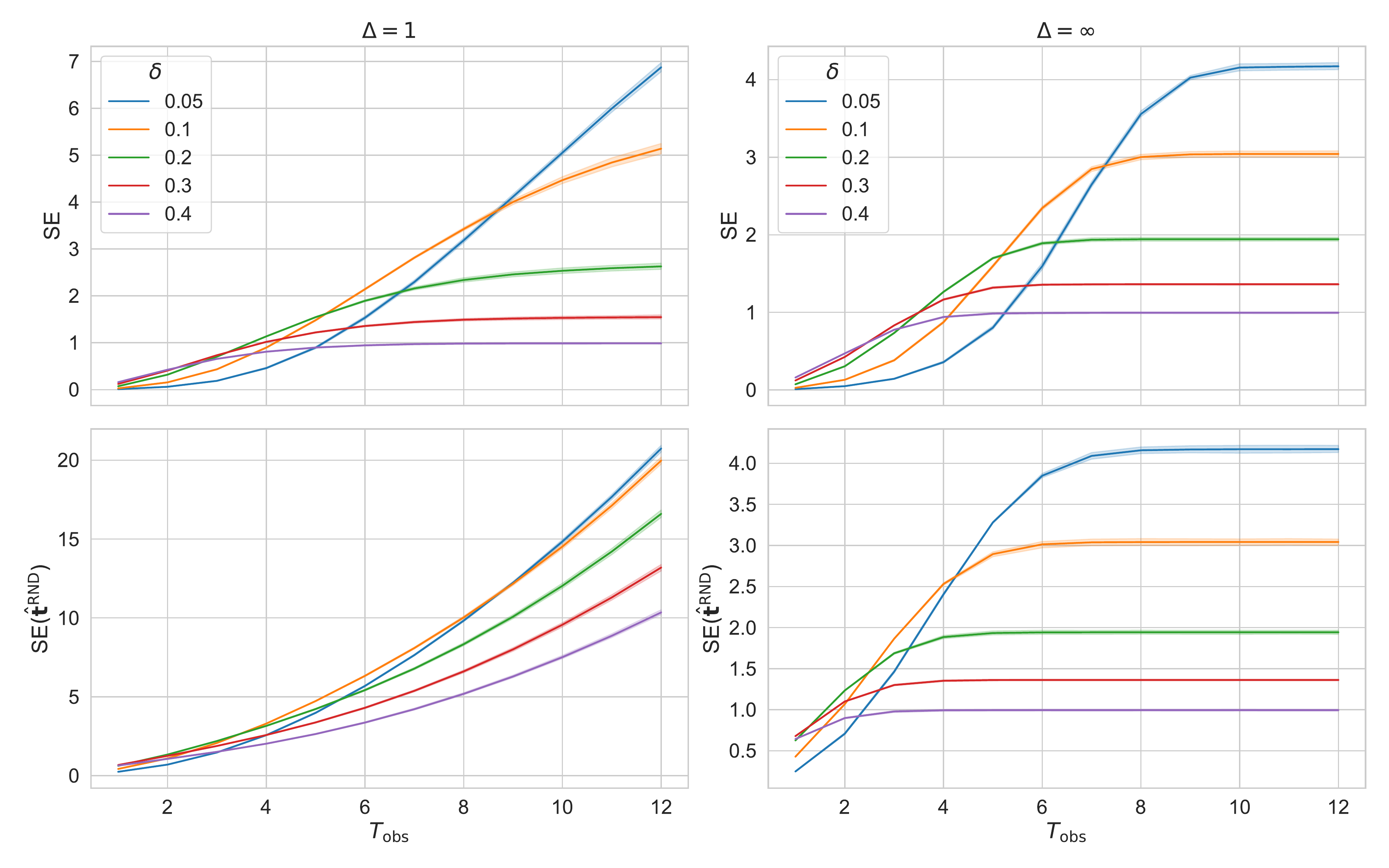}
    \caption{\textbf{Compendium of figure~\ref{fig:PhDiag_RSE_snap}.} Remaining in the same framework as Fig.~\ref{fig:PhDiag_RSE_snap}, we fix $\lambda=0.6$ and study both the SI model and the dSIR model with delay $\Delta=1$. For both scenarios, in the upper panel, we plot the squared error~(\ref{eq:se}) as a function of the snapshot time, and in the lower panel we do the same for the squared error computed with the random estimator defined in Section~\ref{sec:BaySet}.}
    \label{fig:SEvsRND}
\end{figure}
Let's start by looking at the SI model case, displayed in the right panels of Fig.~\ref{fig:SEvsRND}. We can see that both observables, when increasing $T_{\rm obs}$, reach some plateau at some value which depends on the fraction of sources $\delta$. Since the value of the plateau is the same for both estimators, it means that for high values of $T_{\rm obs}$ the rescaled squared error defined in (\ref{eq:RSE}) will go to zero eventually.

Conversely, for the dSIR model, we see that the same behavior is valid for the SE computed with the optimal estimator but not for the one using the random estimator. Indeed, as discussed in the main text, the latter estimator performs quadratically worse when increasing $T_{\rm obs}$ in the case of a fraction of nodes remaining susceptible at the end of the epidemic. This, in turn, makes the rescaled squared error (\ref{eq:RSE}) grow when increasing $T_{\rm obs}$, even if the SE is constant, and finally leads to the non-monotonic behaviors showed in Fig.~\ref{fig:PhDiag_RSE_snap}.
\subsection{Squared error between BP messages}
In this section, we show how the absence of the critical slowing down on the convergence of the BP equations that we have presented in Fig.~\ref{fig:Noslow} can be also seen directly by looking at the BP messages. Specifically, we generate an instance of the problem, i.e. we fix the graph, the sources, the epidemic process and the observations, and then we run the BP equations with \textit{rnd} and \textit{inf} initialization of the  messages. Then, at each iteration time, we compare the two sets of messages by computing the following observable:
\begin{equation}
\label{eq:SEmess}
    {\rm SE}_{\rm mess} = \frac{1}{2|E|(T+2)^2} \sum_{\{i\rightarrow j\}}\sum_{t_i=-1}^T\sum_{t_j=-1}^T (m^{\rm inf}_{i\rightarrow j}(t_i,t_j) - m^{\rm rnd}_{i\rightarrow j}(t_i,t_j))^2
\end{equation}
where the first sum is done over all the ordered pairs in the graph.
\begin{figure}
    \centering
    \includegraphics[width=\textwidth]{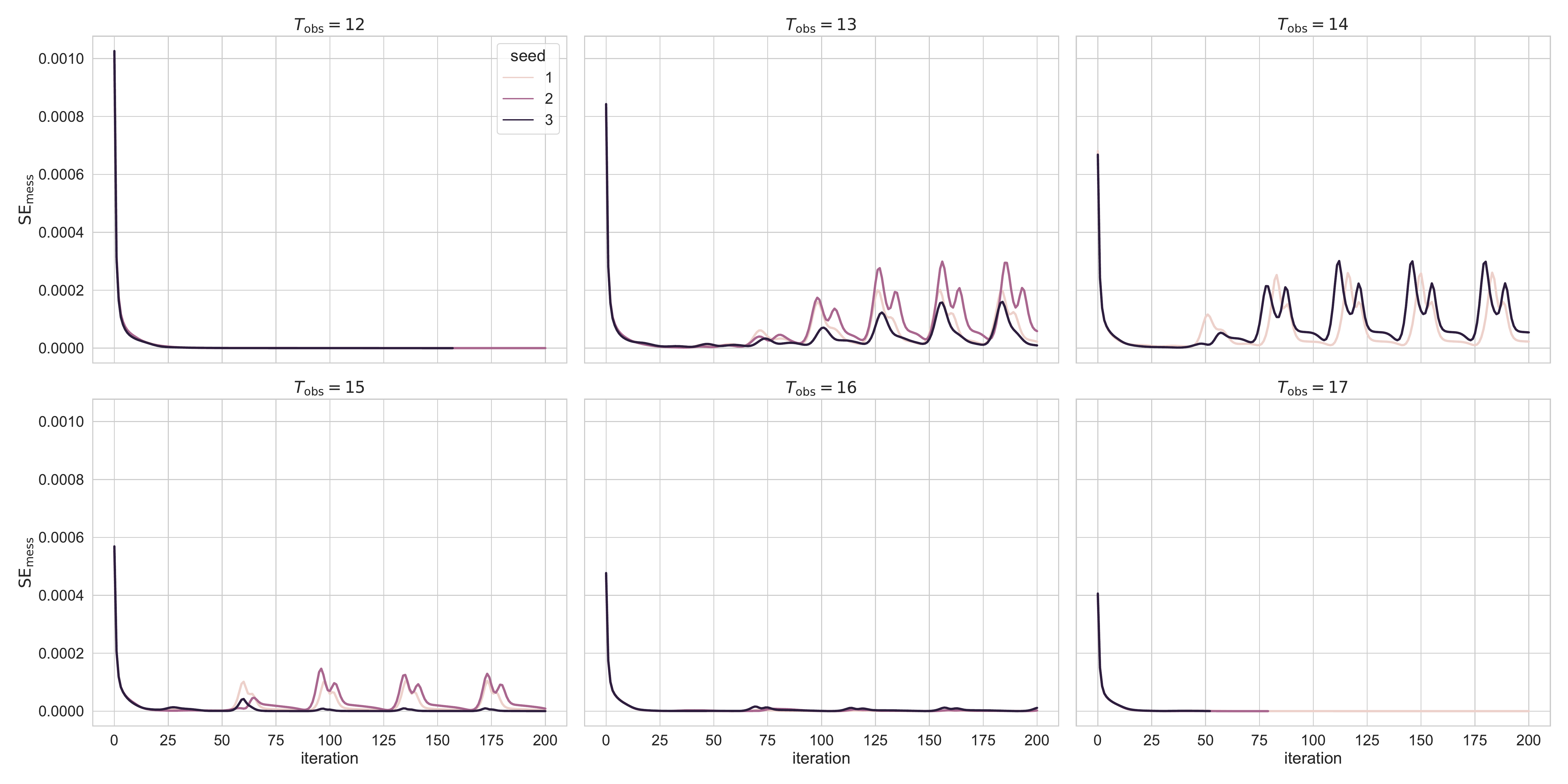}
    \caption{\textbf{Absence of critical slowing down in the convergence of the BP messages.} The plots were generated using an ensemble of 3-regular random graphs (3-RRGs) with $N = 10^4$, $\lambda = 0.5$ and $\delta = 0.005$. We consider observations via a snapshot on an SI model, where the snapshot time is changed in each panel between $T_{\rm obs} = 12$ (top left), and $T_{\rm obs} = 17$ (bottom right). For each panel, we examine three instances of the problem generated randomly, and we monitor the squared error defined in (\ref{eq:SEmess}) during the BP iterations. The damping parameter $\eta$ was fixed to 0.4 for all iterations. We observe that the BP messages of two initializations quickly converge to almost the same values in every case, without a critical slowing down in the non-convergence region, where the oscillations are observed. }
    \label{fig:SEmess}
\end{figure}
The results for the setting of Fig.~\ref{fig:Noslow} are presented in Fig.~\ref{fig:SEmess}. One can clearly see that the behaviour for the first few dozen iterations is the same, with the squared error going quickly to zero, whatever the value of $T_{\rm obs}$. Then, we see that while for high and low values of $T_{\rm obs}$ the squared error remains low and the algorithm converges, for intermediate values at some point it becomes bigger again, and we can see the typical oscillations that characterise the Non-Bayes-Optimal regime.

\clearpage
\bibliographystyle{plain}
\bibliography{citations}

\end{document}